\documentclass[english, reqno]{amsart}
\usepackage{hyperref}
\hypersetup{colorlinks=true,linkcolor=magenta,anchorcolor=green,citecolor=cyan,filecolor=black,menucolor=black,urlcolor=brown}

\usepackage{cite}
\usepackage[table]{xcolor}

\usepackage{mathtools}
\usepackage{enumitem}
\usepackage{cleveref}
\usepackage{mathrsfs}
\usepackage{amsthm}
\usepackage{amscd}
\usepackage{amstext}
\usepackage{amssymb}
\usepackage{graphicx}
\usepackage{setspace}
\usepackage{esint}
\usepackage{rotating}
\usepackage{bm}
\usepackage{nicefrac}
\usepackage{shuffle}
\usepackage{stackrel}
\usepackage{fancybox}
\usepackage{tikz}
\usepackage{tikz-cd}
\usetikzlibrary{shapes.arrows,calc,decorations.pathreplacing,decorations.markings,bending,arrows,positioning}
\tikzset{
    >=stealth',
    punkt/.style={
           rectangle,
           rounded corners,
           draw=black, very thick,
           text width=6.5em,
           minimum height=2em,
           text centered},
    pil/.style={
           ->,
           thick,
           shorten <=2pt,
           shorten >=2pt,}
}

\allowdisplaybreaks

\usepackage{amsfonts}
\usepackage{amssymb}
\usepackage{shuffle}
\usepackage[all]{xy}
\usepackage{array}
\usepackage{lmodern}
\usepackage[OT2,T1]{fontenc}

\usepackage{linearb}

\crefformat{footnote}{#2\footnotemark[#1]#3}

\numberwithin{equation}{section}
\numberwithin{figure}{section}
\makeatother

\usepackage{babel}
\newcount\hh
\newcount\mm
\mm=\time
\hh=\time
\divide\hh by 60
\divide\mm by 60
\multiply\mm by 60
\mm=-\mm
\advance\mm by \time
\def\thistime{\number\hh:\ifnum\mm<10{}0\fi\number\mm}
\def\Li#1(#2){\textrm{Li}_{#1}\left(#2\right)}
\def\cLi_#1(#2){\mathcal{L}_{#1}\left(#2\right)}
\def\bLi_#1(#2){\mathbf{L}_{#1}\left(#2\right)}

\title[Closed and Open string amplitudes]{\bf 
Building blocks of closed and open string amplitudes} 

\author[P. Vanhove]{Pierre Vanhove}
 \address{
 Institut de Physique Th{\'e}orique\\
CEA, IPHT, F-91191 Gif-sur-Yvette, France\\
CNRS, URA 2306, F-91191 Gif-sur-Yvette, France}
\address{
National Research University Higher School of Economics, Russian Federation
}
\author[F. Zerbini]{Federico Zerbini}
 \address{
Institut de Recherche Math\'ematique Avanc\'ee (IRMA)\\
Universit\'e de Strasbourg\\
7 rue Ren\'e Descartes, 67084 Strasbourg, France}
\thanks{IPHT-t20/046}
\date{\today}

\begin{document}

 \begin{abstract}
   In this text we review various relations between building blocks of closed and open
   string amplitudes at tree-level and genus one.  We explain
   that KLT relations between tree-level closed and open string
   amplitudes follow from the holomorphic factorisation of conformal correlation functions on conformal blocks. We give a simple hands-on evaluation of the
   $\alpha'$-expansion of tree-level closed string amplitudes displaying the special single-valued nature of the
   coefficients. We show that the same techniques can be used also at genus-one, where we give a new proof of the single-valued nature of the coefficients of 2-point closed string amplitudes. We conclude by giving an overview of some open problems. (This article is a contribution to the proceedings
   of the workshop ``MathemAmplitudes 2019: Intersection Theory and
   Feynman Integrals'' held in Padova, Italy on 18-20 December 2019.)

\end{abstract}
\maketitle
\newpage
\tableofcontents
\newpage
%%%%%%%%%%%%%%%%%%%%%%%%%%%%%%%%%%%%%%%%%%%%%%%%%%%%%%%%%%%%%%%%%%
\part{Introduction}
\label{sec:introduction}

We are still uncovering remarkable properties of string theory
amplitudes.
Perturbative string theory is characterised by a series expansion with
respect to the string coupling constant $g_s$. At each order in
perturbation theory one can consider the expansion in  the small
inverse string tension $\alpha'$  leading to  quantum field theory scattering together with
an infinite set of stringy quantum corrections.  This  double expansion
can reveal a lot of deep properties of the low-energy effective action of
quantum gravity on various background configurations.

It is remarkable that the 
ultra-violet divergences of maximal supergravity amplitudes in various
dimensions~\cite{Bern:1998ug,Bern:2008pv} are in
perfect agreement with 
the low-energy expansion of string theory amplitudes up to genus three and are fully compatible with the non-perturbative
U-duality symmetries of string theory~\cite{Green:1982sw,Green:2006gt,Green:2006yu,Green:2008bf,Green:2010wi,Green:2010sp,Tourkine:2012ip,Bossard:2015oxa,Pioline:2015yea,Bossard:2017kfv,Pioline:2018pso}.
Recent works have shown a perfect match between this double expansion
and the flat-space limit of correlation functions in $\text{AdS}_5\times S^5$,
giving a remarkable confirmation of the AdS/CFT correspondence~\cite{Alday:2018pdi,Chester:2019jas,Chester:2020dja,MBGzoomamplitude}.

There are five consistent supersymmetric string theories in ten
dimensions composed of four closed superstring theories--the
type~IIA, type~IIB and the two heterotic strings--and the type~I 
superstring theory, which contains also open strings~\cite{Green:1987sp,Polchinski:1998rq,Blumenhagen:2013fgp}.
These string theories are unified by duality symmetries~\cite{Polchinski:1995df,Antoniadis:1996vw} which are the landmark of the special mathematical
and physical properties of string theory.
In this text we are interested in the perturbative expansions of closed and open string amplitudes, and how they are related.

The type~I superstring theory is an orientifold 
of the closed string type~IIB theory obtained by quotienting by the
action of the world-sheet parity reversal $\Omega : (\sigma,\tau)\to
(2\pi -\sigma,\tau)$ (see~\cite{Angelantonj:2002ct}  for a general review of
orientifolds). At each order in perturbation, the type I superstring
amplitude involves a sum of two-dimensional super-surfaces obtained 
from the action of the involution $\mathcal I(z)$, induced by the
orientifold action $\Omega$. The type I superstring theory contains
open and closed strings, possibly non-oriented, which lead to anomaly cancellation~\cite{Green:1984ed,Green:1984qs} and
Ramond-Ramond tadpole cancellation~\cite{Angelantonj:2002ct,Gimon:1996rq}.

Closed (oriented) string amplitudes, which appear in the four closed superstring theories, are computed by integrals over the moduli space
of super-Riemann surfaces of genus~$h$, the number of handles,
with~$n$ punctures and no boundaries. Each order in perturbation
theory is
weighted by the string coupling constant $g_s\ll1$ to the
power of the Euler characteristic of the surface, in this case~$g_s^{2h-2}$.

In type~I superstring theory, involutions $\mathcal I(z)$ act on the (closed) super-Riemann
surfaces to remove $b$
discs and add $c$ cross-caps (obtained by
gluing a real projective plane) to the surface. The
loci of fixed points of the involution are the boundaries of the  world-sheet, which contain the open string insertions. 
Each order in perturbation is
weighted by the string coupling constant $g_s\ll1$ to the
power of the Euler characteristic of the surface, in this case $g_s^{2h+b+c-2}$. 

On a surface $M$ obtained by applying an involution $\mathcal I_M(z)$ to a closed (oriented) world-sheet, the Green function is given in terms of the closed string Green function by the formula~\cite{Burgess:1986ah,Burgess:1986wt}:
\begin{equation}\label{e:Gimage}
  G^{M}(z,z')= \frac12 \left( G^{\rm closed}(z,z')+ G^{\rm
  closed}(\mathcal I_M(z), z')+    G^{\rm
  closed}(z,\mathcal I_M(z'))+ G^{\rm
  closed}(\mathcal I_M(z),\mathcal I_M( z'))\right)\,.
\end{equation}

At tree-level, the Kawai, Lewellen and Tye (KLT) relations~\cite{Kawai:1985xq} give
a construction of the closed string integrals on the sphere with $n$
punctures as a sum of products of pairs of open string amplitudes
with $n$ marked points on the boundary of the disc.
This relation has led to the fundamental perturbative  duality
between colour factors of gauge theory and kinematic factors in field
theory and string theory~\cite{Bern:2008qj,Bern:2019prr}.

Hints that similar relations may generalise at genus one come from the small
$\alpha'$ expansion of one-loop open and closed string
amplitudes~\cite{Broedel:2018izr,Gerken:2018jrq,Mafra:2019ddf,Mafra:2019xms,Zagier:2019eus,Gerken:2020yii}.  Much less
is known at higher-loop orders but one can reasonably suspect that
similar relations exist as being a consequence of the intrinsic
relation between closed and open string amplitudes. 

With the aim of finding a higher-loop generalisation of the KLT
relations in string theory, we have shown in~\cite{Vanhove:2018elu} that the relation between  closed string  tree-level
amplitudes and open string tree-level amplitudes is a particular application of
the conformal block decomposition of correlation functions  in
two-dimensional minimal model conformal field theory. This construction is generalizable at higher
genus~\cite{DiFrancesco:1997nk,DHoker:1989cxq}. The  chiral splitting of the closed string conformal blocks at fixed internal
momenta gives a realisation of the double-copy of open string
integrands before integration over the loop momentum
(see~\cite{Mafra:2018qqe} for a discussion at genus one). 
Therefore, this approach may
lead to a higher-loop generalisation of the KLT relations in string theory. 

The conformal block approach, and  its restriction to amplitudes given
by the KLT relation, inverts the involution construction for tree-level amplitudes,
as illustrated on the following diagram:

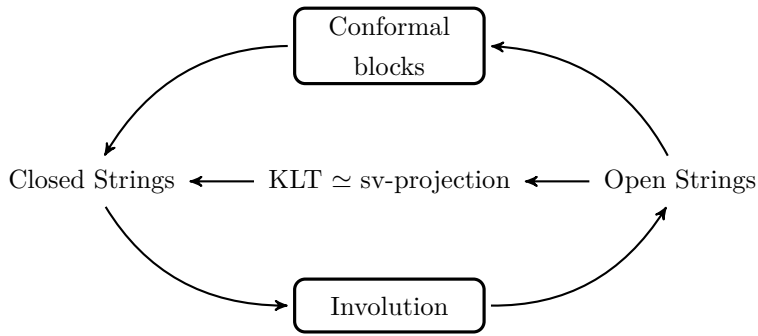
\begin{center}\begin{tikzpicture}[node distance=1cm, auto,]
   \node[] (svmap) {KLT $\simeq$  sv-projection};
   \node[punkt,above=of svmap] (cft) {Conformal blocks};
      \node[punkt,below=of svmap] (projection) {Involution};
       \node[right=of svmap] (t) {Open Strings}
        edge[pil] (svmap.east)
 edge[pil,<-,bend left=30] (projection.east)
   edge[pil,bend right=30] (cft.east); 
   \node[left=of svmap] (g) {Closed Strings}
     edge[pil,<-] (svmap.west)
 edge[pil, bend right=30] (projection.west)
 edge[pil,<-, bend left=30] (cft.west);
 \end{tikzpicture}
\end{center}

This article is divided into three parts, and the introduction
constitutes the first one. Part~\ref{sec:genus-0} is about the
building blocks of tree-level perturbative string amplitudes. We start
by reviewing the result~\cite{Klebanov:1995ni, Hashimoto:1996kf, Garousi:1996ad,Garousi:2006zh,Stieberger:2009hq,Chen:2009tr,Aldi:2020dvw} that all tree-level (bosonic or type I super-)
string amplitudes on the disc (with both open
and closed string states) and on the real projective plane (where only
closed string states appear) are combinations of open string building
blocks given by ordered integrals on the boundary of the disc. In a
similar way, tree-level closed (bosonic, type~IIA or type~IIB or
heterotic) string amplitudes are combinations of closed string
building blocks. All tree-level (bosonic or super-) string amplitudes
are obtained from appropriate choices of the kinematic coefficients in
the linear combinations of the building blocks.

We then give the relation between the open
string and closed string building blocks using the conformal block
decomposition approach of~\cite{Vanhove:2018elu}. The open string
building blocks are special values of the (multi-valued) conformal blocks
and the closed string building blocks are special values
of (single-valued) conformal correlation functions. 
The modern formulation of the KLT relations in terms of twisted cohomology theory as
twisted bilinear period relations~\cite{Mizera:2017cqs, Brown:2018omk} should generalise and give in this context a cohomological interpretation of the conformal block decomposition.

The cohomological viewpoint can be used to relate closed string amplitudes to the newborn theory of single-valued periods~\cite{BrownSVMZV, BrownDupont}, providing a mathematical interpretation of the fact that open and closed string amplitude building blocks are special values of multi-valued and single-valued functions, respectively. This explains why the $\alpha'$-expansion of the closed string building blocks involves only a very small subspace of multiple zeta values, given precisely by those called ``single-valued'' which come from the theory of single-valued periods~\cite{Stieberger:2009rr, Stieberger:2013wea, Stieberger:2014hba, Fan:2017uqy, Schlotterer:2018zce, Brown:2018omk, Vanhove:2018elu}. Following~\cite{Vanhove:2018elu}, we illustrate this phenomenon by giving a simple hands-on evaluation of the
$\alpha'$-expansion of closed string amplitudes based on the ``single-valued
integration'' of (single-valued) multiple polylogarithms developed in~\cite{Schnetz:2013hqa, DelDuca:2016lad, Vanhove:2018elu}. 

Part~\ref{sec:genus-one} is about the building blocks of one-loop perturbative string amplitudes. On the one hand, we have closed (oriented) superstring amplitudes on the torus, arising from the four closed superstring theories. They can be expanded on closed string building blocks as in the genus zero case. The low-energy expansion of these building blocks gives rise to modular graph functions and forms~\cite{DHoker:2015wxz,DHoker:2018mys,DG16,Gerken:2018jrq}, which will be reviewed in section~\ref{sec:modul-graph-funct} and which are conjecturally related to the same single-valued periods appearing at genus zero. In particular, in section~\ref{sec:two-points-function} we present a new proof of a special case of this conjecture, adapting to this setting the single-valued integration method already used at tree level. Even though this is not a new result (two different proofs have already appeared in the recent papers~\cite{DHoker:2019xef,Zagier:2019eus}), our new approach may be of interest because it seems more likely to generalize.

On the other hand, we have type~I superstring amplitudes given by applying involutions on the torus. One finds superstring amplitudes on the annulus (with both open and closed string states), on the M\"obius strip (unoriented, with both open and closed string states) and on the Klein bottle (unoriented, only closed string states appear). The low-energy expansion of amplitudes with open string
states on the boundaries of the annulus or the M\"obius strip involves elliptic analogues of multiple zeta values~\cite{Broedel:2014vla,Broedel:2015hia, BMRS}. In this case, it was observed in~\cite{Broedel:2018izr} that open and closed string building blocks seem to be related by a genus-one analogue of the single-valued projection from genus-zero. This points towards the possibility of extending KLT relations to genus one. In section~\ref{sec:esv} we briefly review the simplest instance of this relation, which was demonstrated in~\cite{Zagier:2019eus}. 

Finally, very little is known about the low-energy expansion of one-loop string amplitudes with both open and closed string states. For instance, we do not know if it is possible to reduce them to open string building blocks (annulus amplitudes with boundary insertions only), as in the tree-level case. This is one of the open questions that we list in section~\ref{sec:open}, which concludes the paper.

%%%%%%%%%%%%%%%%%%%%%%%%%%%%%%%%%%%%%%%%%%%%%%%%%%%%%%%%%%%%%%%%%
\section*{Acknowledgments}
%%%%%%%%%%%%%%%%%%%%%%%%%%%%%%%%%%%%%%%%%%%%%%%%%%%%%%%%%%%%%%%%
We would like to thank  C. Angelantonj, O. Schlotterer, P. Tourkine
for discussions and comments on this text. 
P. Vanhove thanks the organisers of the workshop ``MathemAmplitudes 2019:
Intersection Theory and Feynman Integrals'' held in Padova, Italy on
18-20 December 2019, for giving the opportunity to present some of the
results reported in this text. The research of P. Vanhove has received funding from the ANR grant
  ``New Structures in Amplitudes'' ANR-17- CE31-0001-01, and is partially supported by
  Laboratory of Mirror Symmetry NRU HSE, RF Government grant,
  ag. N$^\circ$ 14.641.31.0001. The research of F. Zerbini was supported by the LabEx IRMIA.

  \part{Tree-level amplitudes}
\label{sec:genus-0}

\section{Building blocks of closed and open string amplitudes}
\label{sec:build-blocks-clos}

\subsection{Closed oriented tree-level amplitudes}
\label{sec:closed-string-tree}

Any tree-level $N+3$-point closed oriented string amplitude ($N\geq 1$) can be written as a finite linear combination  of partial amplitudes~\cite{Polchinski:1998rq,Green:1987sp,DHoker:1988pdl}
\begin{equation}\label{Firsteq}
  M_{N+3}(\pmb s,\pmb \epsilon)=\sum_r c_r(\pmb s,\pmb \epsilon) M_{N+3}(\pmb
  s,\pmb n^r,\pmb {\tilde n}^r)\,.
\end{equation}
The coefficients $c_r(\pmb s,\pmb \epsilon)$ are rational functions of
the kinematic variables $\pmb s=(s_{ij})_{1\leq i< j\leq N+3}:=(- {1\over2} k_i\cdot k_j)_{1\leq i< j\leq N+3}$ (where $k_i$ are
external momenta, subject to momentum conservation
$k_1+\cdots +k_{N+3}=0$ and on-shell condition $\alpha' k_i^2\in
4\mathbb Z$ lower or equal to $4$ with the metric convention $(-,+,\dots,+)$), the polarisation tensors $\pmb
\epsilon=(\epsilon_i)_{1\leq i\leq N+3}$ and the
colour factors for the heterotic string amplitudes. These kinematic
coefficients determine the specifics of the
closed string theory (bosonic string, superstring or heterotic string)
or of the external states (polarisation tensors and colour factors). The partial amplitudes $M_N(\pmb
s,\pmb n,\pmb {\tilde n})$, defined for integers $\pmb n=(n_{ij})_{1\leq i< j\leq N+3}$, $\pmb{\tilde{n}}=(\tilde{n}_{ij})_{1\leq i< j\leq N+3}$ and implicitly depending also on the inverse string tension~$\alpha'$, are integrals on~$N$ copies of the Riemann sphere~$\mathbb{P}^1_\mathbb{C}$ of the form 
\begin{multline}\label{e:partamplitudesG}
  M_{N+3}(\pmb s,\pmb n,\pmb{\tilde n})\,= \\
  =\,\int_{(\mathbb{P}^1_{\mathbb{C}})^{N}} \prod_{i=1}^{N} d^2w_i \prod_{1\leq i<j\leq N+2}
 e^{2\alpha' s_{ij} G^{P^1_\mathbb C}(w_i,w_j) }
 (\partial_{w_j} G^{P^1_\mathbb C}(w_i,w_j))^{-2n_{ij}}
 (\partial_{\bar w_j} G^{P^1_\mathbb C}(w_i,w_j))^{-2\tilde n_{ij}}\,,
  \end{multline}
where $w_{N+1}:=0$, $w_{N+2}:=1$ and $d^2w_i:=\tfrac{idw_id\bar{w}_i}{2\pi}$. The Green function on the Riemann sphere is
 \begin{equation}
   G^{P^1_\mathbb C}(z,z'):=-\frac12\log|z-z'|^2\,,
 \end{equation}
 so that the partial amplitudes reads 
 \begin{multline}\label{e:partamplitudes}
  M_{N+3}(\pmb s,\pmb n,\pmb{\tilde n})= \int_{(\mathbb{P}^1_{\mathbb{C}})^{N}} \prod_{i=1}^{N} d^2w_i \prod_{1\leq i<j\leq N+2}
 |w_i-w_j|^{-2\alpha' s_{ij} }
 (w_i-w_j)^{n_{ij} }  (\bar w_i-\bar
 w_j)^{\tilde n_{ij}}\,.
  \end{multline}
For instance, the Virasoro-Shapiro amplitude is a particular case
with $N=1$ and vanishing integers~$n_{ij}, \tilde{n}_{ij}$:
\begin{equation}\label{e:VirasoroShapiro}
   M_4(\pmb s,\pmb 0,\pmb 0)=\int_{\mathbb{P}^1_{\mathbb{C}}} |w|^{-2\alpha's_{12} }
                           |1-w|^{-2\alpha's_{13}}\,d^2w
={\Gamma( -\alpha's_{12}+1) \Gamma(-\alpha'
    s_{13} +1)\Gamma(-\alpha's_{23} -1)\over \Gamma(\alpha' s_{12}
   )\Gamma(\alpha's_{13})\Gamma(\alpha's_{23}+2)}\,.
 \end{equation}

\subsection{Open and unoriented tree-level amplitudes}
\label{sec:open-string-tree}
At tree level, open oriented (super)string amplitudes are defined over the disc and closed unoriented (super)string amplitudes are defined over the real projective plane.

\begin{figure}[h]
  \centering
\begin{tikzpicture}
  \shade[ball color = gray!40, opacity = 0.4] (0,0) circle (2cm);
  \draw (0,0) circle (2cm);
  \draw (-2,0) arc (180:360:2 and 0.6);
  \draw[dashed] (2,0) arc (0:180:2 and 0.6);
  \fill[fill=black] (0,0) circle (1pt);
  % \draw[dashed] (0,0 ) -- node[above]{$r$} (2,0);
  \draw (0,-2.5) node{(a)};
\end{tikzpicture}
\qquad
\begin{tikzpicture}
 \shade[ball color = gray!40, opacity = 0.4] (0,0) circle (2cm);
  \draw (0,0) circle (2cm);
  \draw (-2,0) arc (180:360:2 and 0.6);
  \draw[dashed] (2,0) arc (0:180:2 and 0.6);
  \fill[fill=black] (0,0) circle (1pt);
  % \draw[dashed] (0,0 ) -- node[above]{$r$} (2,0);
  \fill[fill=red] (1,1) circle (1pt);
  \fill[fill=red] (1,-1) circle (1pt);
  \draw[dashed,color=red] (1,-1 ) --(1,1);
  \draw[color=red] (1,1) node[left]{$P$};
  \draw[color=red] (1,-1) node[left]{$P'$};
    \draw (0,-2.5) node{(b)};
  \end{tikzpicture}
  \qquad
\begin{tikzpicture}
 \shade[ball color = gray!40, opacity = 0.4] (0,0) circle (2cm);
  \draw (0,0) circle (2cm);
  \draw (-2,0) arc (180:360:2 and 0.6);
  \draw[dashed] (2,0) arc (0:180:2 and 0.6);
  \fill[fill=black] (0,0) circle (1pt);
  \fill[fill=red] (1,1) circle (1pt);
  \fill[fill=red] (-1,-1) circle (1pt);
   \draw[dashed,color=red] (-1,-1 ) --(1,1);
   % \draw[dashed] (0,0 ) -- node[above]{$r$} (2,0);
    \draw[color=red] (1,1) node[left]{$P$};
    \draw[color=red] (-1,-1) node[left]{$P'$};
      \draw (0,-2.5) node{(c)};
\end{tikzpicture}
\caption{$(a)$ The Riemann sphere, $(b)$ the disc and $(c)$ the real
  projective plane. The disc and the real projective plane are obtained from the action of an
  involution identifying the points $P$ and $P'$ on the Riemann sphere.}
\label{fig:tree}\end{figure}
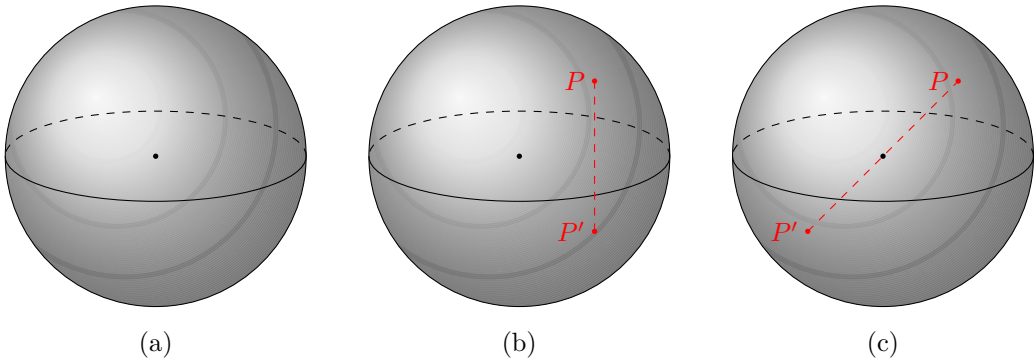

\subsubsection{The disc amplitudes}
\label{sec:disc}
The disc $\mathbb{D}_2$ is obtained by the action of the involution $\mathcal
 I^{\mathbb{D}_2}(z)=\bar z$ on the Riemann sphere: $\mathbb{D}_2\cong \mathbb P^1_\mathbb C/\mathcal
 I_{\mathbb{D}_2}$ (see figure~\ref{fig:tree}(b)).
The  bosonic string propagator on the disc $\mathbb{D}_2$ is obtained from the method of images by applying equation~\eqref{e:Gimage}, and reads
\begin{equation}\label{e:Gdiscbulk}
  G^{\mathbb{D}_2}(z,z')=-\frac12(\log|z-z'|^2
  +  \log|\bar z-z'|^2).
\end{equation}
On the boundary of the disc where $z=x\in\mathbb R$ and
$z'=y\in\mathbb R$ we have 
\begin{equation}\label{e:Gdiscbdry}
  G^{\mathbb{D}_2}(x,y)= -2 \log|x-y|\,.
\end{equation}
One can have all vertex operators inserted on the boundary
of the disc. For instance, using the Green function above, identifying the boundary of the disc with $\mathbb{R}$ and fixing by conformal invariance three tachyon states at $0,1,\infty$, one obtains the celebrated Veneziano amplitude
\begin{equation}\label{e:VenezianoT}
  \int_0^1 x^{-4\alpha's_{12}}(1-x)^{-4\alpha' s_{13}} dx\,=\,
  {\Gamma(-4\alpha's_{12}+1)\Gamma(- 4\alpha's_{13}+1)\over \Gamma(2-4\alpha's_{12}-4\alpha's_{13})}\,.
\end{equation}
More generally, four-point amplitudes of (open) bosonic strings involving both tachyons and massless states on the boundary of the disc differ from eq.~(\ref{e:VenezianoT}) by a proportionality factor which is a rational function in the kinematic variables $\alpha' s_{12}$, $\alpha' s_{13}$.

One can also consider mixed amplitudes with insertion of $N_o$ open string external states on the boundary of the
disc and $N_c$ closed string external  states in the bulk of the disc.
Such amplitudes were shown to  be reducible to linear combinations of
disc amplitudes with $N_o+2N_c$ open string states on the boundary of the
disc~\cite{Klebanov:1995ni, Hashimoto:1996kf,Garousi:1996ad,Stieberger:2009hq}.

As an illustration we present the case of the insertion of
two states of momenta $k_1$ and $k_2$ at the respective position $z_1$
and $z_2$ in the bulk of the disc. The method of image on the
covering plane doubles the insertions into  two
insertions at the positions $z_1$ and $\bar z_1$
and two insertions  at the positions $z_2$ and $\bar z_2$.
By conformal invariance we fix $z_1=0$ and $z_2=ix$ with
$x\in[0,1]$. The integration
over $z_1$  and $z_2$ is reduced to a single integration over
$x\in[0,1]$. Using the Green function in~\eqref{e:Gdiscbulk}
the two-point amplitude is then given by an integral of the type 
\begin{equation}\label{e:D2grav}
  \int_0^1   (x^2)^{-4\alpha's_{12}-1} (1-x^2)^{-2\alpha's_{13}}
    dx^2
= -2\alpha's_{13} {\Gamma(-2\alpha's_{12})\Gamma(-2\alpha's_{13})\over
  \Gamma(1-2\alpha' s_{12}-2\alpha's_{13})}\,,
 \end{equation}
where $s_{12}=-\frac12 k_1\cdot k_2$ and $s_{13}=-\frac12 k_1\cdot D\cdot
k_1$ and where $k_i$ is the momentum of the external state and  $D\cdot
k_i$ is the momentum of its image.  
Using the functional equation $\Gamma(x+1)=x\Gamma(x)$, we recognise
that this amplitude only differs from the Veneziano amplitude
in~\eqref{e:VenezianoT} just by a proportionality factor which is a rational function of the kinematic invariants (up to scaling the variables).

\subsubsection{The real projective plane amplitudes}
\label{sec:real-proj-plane}
The action of the involution $\mathcal I_{\mathbb
   P_{\mathbb{R}}^2}(z)=-1/\bar z$ on  the Riemann sphere gives the real projective plane $\mathbb
   P_{\mathbb{R}}^2\cong \mathbb P^1_\mathbb C/\mathcal I_{\mathbb
   P_{\mathbb{R}}^2}$ (see figure~\ref{fig:tree}(c)).
The  bosonic string propagator on the real projective plane
$\mathbb P_{\mathbb{R}}^2$ is obtained from the method
of images by applying equation~\eqref{e:Gimage}, and reads
\begin{equation}\label{e:Gprojective}
  G^{\mathbb P_{\mathbb{R}}^2}(z,z')=-\frac12\left(\log|z-z'|^2
  +\log|1+z'\bar z|^2-\log|zz'|^2\right)\,.
\end{equation}
The real projective plane does not have boundaries, so the amplitudes
only involve closed string states. The amplitudes with $N_c$ closed string state insertions are, as in the previous section, reducible to linear combinations of disc amplitudes with $2N_c$ open string states inserted on the boundary of the disc~\cite{Garousi:2006zh,Chen:2009tr,Aldi:2020dvw}.  
For instance, for the case of an amplitude with the
insertion of  two closed string states, the method of images on the
covering plane doubles the insertions into  two
insertions at the positions $z_1$ and $\bar z_1$
and two insertions  at the positions $z_2$ and $\bar z_2$.
By conformal invariance we fix $z_1=0$ and $z_2=ix$ with
$x\in[0,1]$. The integration
over $z_1$  and $z_2$ is reduced to a single integration over
$x\in[0,1]$. Using the Green function in~\eqref{e:Gprojective}
the two-point function is then given by an integral of the type
\begin{equation}\label{e:RP2grav}
  \int_0^1  \left( (x^2)^{-4\alpha's_{12}-1} + (x^2)^{-4\alpha's_{12}-4\alpha's_{13}-1} \right)  (1-x^2)^{-2\alpha's_{13}}
    dx^2
=- 2\alpha's_{13} {\Gamma(-2\alpha's_{12})\Gamma(2\alpha's_{13}+2\alpha's_{13})\over
  \Gamma(1+2\alpha' s_{13})}
 \end{equation}
where $s_{12}=-\frac12 k_1\cdot k_2$ and $s_{13}=-\frac12 k_1\cdot D\cdot
k_1$
This  amplitude differs from the Veneziano
amplitude in~\eqref{e:VenezianoT} only by a proportionality factor which is a rational function of the
kinematic invariants (up to scaling the variables).

\medskip
We see on these examples that the main properties of the amplitude are not
dictated by the local structure of the Green functions on the
world-sheet, but by the boundary conditions on the integration space
imposed by the action of the involution used to construct the open
string geometry.

\bigskip 
To summarize, all amplitudes on the disc or the real projective plane are reducible to open string amplitudes on the boundary of a disc
\begin{equation}\label{e:Acolor}
  A_{N+3}(\pmb s,\pmb \epsilon)=\sum_{r,\rho} d_{r,\rho}(\pmb s,\pmb \epsilon)
  A_{N+3}(\pmb s,\pmb n^r,\rho)
\end{equation}
expanded on a basis of ordered integrals $A_{N+3}(\pmb s,\pmb n^r,\rho)$ along the real line, indexed by permutations $\rho\in\mathfrak{S}_{N}$, which we refer to as tree-level open string building blocks and we fix to be
\begin{multline}\label{Abuildblock}
  A_{N+3}(\pmb s,\pmb n, \rho)=\int_{x_{\rho(N)}\leq \cdots\leq
  x_{\rho(1)} \leq 0}
 \prod_{1\leq m<n\leq N}
(x_{\rho( n)}-x_{\rho( m)})^{ -4 \alpha' s_{\rho(n)\rho(m)}+n_{\rho(n)\rho(m)}}\cr
\times\prod_{m=1}^{N} (-x_{\rho(m)})^{ -4 \alpha' s_{N+1\rho(m)}+n_{N+1\rho(m)}} (1-x_{\rho(m)})^{-4 \alpha' s_{N+2\rho(m)}+n_{N+2\rho(m)}}\prod_{i=1}^{N} dx_i\,.
\end{multline}

Understanding the  properties of any  tree-level amplitudes in string
theory is reduced to understand the building
blocks~\eqref{e:partamplitudes} and~\eqref{Abuildblock} and their relations.
This is the topic of the next two sections.

\section{Closed string amplitudes from single-valued correlation functions}
\label{sec:conf-block-decomp}

In this section we review the result of\cite{Vanhove:2018elu} that the KLT relation between the closed
string building blocks in~\eqref{e:partamplitudes} and the open string
building blocks in~\eqref{e:Acolor} is a special case of the conformal
block decomposition of string theory correlation
functions.

We introduce, for $\eta\in \mathbb C$ and $N\in\mathbb{N}$, the functions 
\begin{multline}\label{e:Gdefgeneral}
  \mathcal G_N\left({\pmb a\, \pmb b\, \pmb c\, \pmb d\atop \pmb{\tilde a}\,
      \pmb{\tilde  b}\, \pmb{\tilde c}\, \pmb{\tilde d}}\Bigg|\eta\right):=\cr \int_{(\mathbb{P}^1_{\mathbb{C}})^N}  
\prod_{i=1}^{N}  z_i^{a_i} (1-z_i)^{b_i} (\eta-z_i)^{c_i}
\prod_{i=1}^{N}  
 \bar   z_i^{\tilde a_i} (1-\bar z_i)^{\tilde b_i} (\bar \eta-\bar z_i)^{\tilde
   c_i} \prod_{1\leq i<j\leq N}  (z_i-z_j)^{ d_{ij}}   (\bar z_i-\bar z_j)^{\tilde d_{ij}}  
  \prod_{i=1}^{N} d^2z_i \,,
\end{multline}
where the tuples of exponents $\pmb a, \pmb b, \pmb c, \pmb
d,\pmb{\tilde a},\pmb{\tilde  b}, \pmb{\tilde c}, \pmb{\tilde d}$ are
formed by complex numbers which satisfy
\begin{equation}
\begin{aligned}
  \label{e:spinG}
  a_i-\tilde a_i,  b_i-\tilde b_i,  c_i-\tilde c_i,&\in\mathbb Z \qquad &1\leq i\leq N,\cr
d_{ij}- \tilde d_{ij}&\in\mathbb Z\qquad &1\leq i<j\leq N\,.
\end{aligned}
\end{equation}
These  conditions imply that the integrand of~(\ref{e:Gdefgeneral}) is single-valued and therefore that the
  integral over $\mathbb{C}^n$ makes sense and defines a single-valued function of $\eta\in\mathbb{C}\setminus\{0,1\}$ for those exponents such that the integral converges absolutely.

The  partial amplitudes~\eqref{e:partamplitudes} are the values at
$\eta=1$ of the functions~\eqref{e:Gdefgeneral} when the exponents
satisfy the relations
\begin{equation}\begin{aligned}\label{e:paramA}
  a_i&:= -\alpha'   s_{i\,N+1}+n_{i\, N+1},  \quad   b_i+c_i:= -\alpha'  s_{i\, N+2} +n_{i\,N+2},  \quad 1\leq i\leq N,\\
 \tilde a_i&:= -\alpha' s_{i\,N+1}+\tilde n_{i\, N+1},  \quad   \tilde b_i+\tilde
 c_i:=  -\alpha' s_{i\,
   N+2} +\tilde n_{i\,N+2},  \quad 1\leq i\leq N,\\
d_{ij}&:=  -\alpha' s_{ij}  +n_{ij},  \quad \tilde d_{ij}= -\alpha' s_{ij}+\tilde n_{ij}, \qquad\qquad\qquad\qquad\  1\leq i<j\leq
                 N.
\end{aligned}
\end{equation}
  
Following the standard rules for the correlation functions of
conformal field theory minimal models~\cite{DiFrancesco:1997nk} and
string theory
in~\cite{DHoker:1988pdl,Green:1987sp,Polchinski:1998rq,Polchinski:1998rr},
one can express the function $\mathcal G_N$ as a
conformal correlation function:
\begin{equation}\label{e:Gcft}
   \mathcal G_N\left({\pmb a\, \pmb b\, \pmb c\, \pmb d\atop \pmb{\tilde a}\,
      \pmb{\tilde  b}\, \pmb{\tilde c}\, \pmb{\tilde d}}\Bigg|\eta\right)=\int_{(\mathbb{P}^1_{\mathbb{C}})^N} \prod_{i=1}^{N} d^2z_i
  \left\langle V_{0}(0) \prod_{i=1}^{N} V_i(z_i)
  V_{N+1}(1) V_{N+2}(\infty)  U(\eta)\right\rangle\,,
\end{equation}
with $U(\eta)=\,:\exp(i k_*\cdot X(\eta)):$ an auxiliary vertex
operator. The fact that $\mathcal{G}_N$ is single-valued is coherent with the axioms of conformal field theory, because it guarantees the locality of the theory.

\subsection{Holomorphic factorisation}
\label{sec:holom-fact}

It was shown in~\cite{Vanhove:2018elu} that the correlation function~\eqref{e:Gdefgeneral} has the holomorphic factorisation 

\begin{equation}\label{e:GholoFac}
      \mathcal G_{N}\left({\pmb a\, \pmb b\, \pmb c\, \pmb d\atop \pmb{\tilde a}\,
      \pmb{\tilde  b}\, \pmb{\tilde c}\, \pmb{\tilde
        d}}\Bigg|\eta\right)=\frac 1{\pi^N}\,\vec  J_N(\tilde{\pmb
    a},\tilde{\pmb b},\tilde{\pmb c};\tilde{\pmb d};\bar \eta)^T  \,
  \hat G_N\left(\pmb a, \pmb b, \pmb c; \pmb d\right)  \, \vec J_N(\pmb a,\pmb
b,\pmb c;\pmb d;\eta) \,,
\end{equation}
with
\begin{itemize}
\item[(i)] The vector  $\vec J_N(\pmb a,\pmb b,\pmb c;\pmb
d;\eta):=\{J_{(\sigma,\rho)}(\pmb a,\pmb b,\pmb c;\pmb d;\eta),
(\rho,\sigma)\in\mathfrak S_r\times \mathfrak S_s, r+s=N\}$ of  {\sl  Aomoto-Gel'fand hypergeometric functions}~\cite{Aomoto, Gelfand, GelRus}, i.e. the iterated integrals
  \begin{multline}\label{e:IsigmarhoDefdual}
J_{(\rho
  ,\sigma)}(\pmb a,\pmb b,\pmb c;\pmb
                d;\eta)=\int_{\tilde\Delta_{(\rho,\sigma)}(\eta)}
 \prod_{m=1}^r \prod_{n=1}^s (z_{ \sigma(n)}-z_{ \rho(m)})^{d_{ \rho(m) \sigma(n)}} \prod_{1\leq m<n\leq r}
(z_{\rho( m)}-z_{\rho( n)})^{d_{\rho( m) \rho( n)}}\cr
\times
\prod_{1\leq m<n\leq s}
(z_{\sigma(m)}-z_{\sigma(n)})^{d_{\sigma( m) \sigma( n)}}
\prod_{m=1}^{r} (-z_{\rho(m)})^{a_{\rho(m)}} (1-z_{\rho(m)})^{b_{\rho(m)}}(\eta-z_{\rho(m)})^{c_{\rho(m)}}\cr
\times \prod_{n=1}^{s} (z_{\sigma(n)})^{a_\sigma(n)}
  (1-z_{\sigma(n)})^{b_{\sigma(n)}}(z_{\sigma(n)}-\eta)^{c_{\sigma(n)}}\,\prod_{i=1}^{N} dz_i\,,
\end{multline}
integrated over the domain
\begin{equation}
  \label{e:DeltaDualDef}
  \tilde\Delta_{(\rho,\sigma)}(\eta):= \{z_{\rho(r)}\leq \cdots\leq
  z_{\rho(1)} \leq 0\leq \eta\leq z_{\sigma(s)} \leq \cdots
  \leq z_{\sigma(1)}\leq 1 \}\,.
\end{equation}
with $(\rho,\sigma)\in\mathfrak S_r\times \mathfrak S_s$  and
$(\rho,\sigma)\in\mathfrak S_r\times \mathfrak S_k$ the set of
permutation of $k$-elements. 

\item[(ii)] $\hat G_N(\pmb a,\pmb b,\pmb c;\pmb
  d)$
is an invertible  square matrix  of size $(N+1)!$, independent of $\eta$, whose entries are rational linear combinations of exponentials of $\mathbb{Q}(\pi i)$-linear
 combinations of the parameters $a_i$, $b_i$, $c_i$ and $d_{ij}$.
\end{itemize}

The relation~\eqref{e:GholoFac} is the usual holomorphic factorisation
of tree-level conformal correlation function on conformal blocks. The
conformal blocks are the mutivalued
integrals~\eqref{e:IsigmarhoDefdual}. 

\subsubsection{The $N=1$ case}
\label{sec:n=1-case}

We illustrate how the factorisation works in the $N=1$ case
corresponding to the four-point amplitude.
We want to perform the holomorphic factorisation of the integral
  \begin{equation}
    \mathcal G_1 \left({ a\,  b\, c\atop {\tilde a}\,
      {\tilde  b}\, {\tilde c}}\Bigg|\eta\right)=\int_{\mathbb{P}^1_{\mathbb{C}}}   z^{a}
   \bar z^{\tilde a}
     (z-1)^{b}
     (\bar z-1)^{\tilde b}
      (z-\eta)^{c}
      (\bar z-\bar \eta)^{\tilde c}d^2z \,,
    \end{equation}
with the integer spin  conditions
\begin{align}\label{e:spinN1}
  a-\tilde a\in\mathbb Z,\qquad
   b-\tilde b\in\mathbb Z,\qquad
c-\tilde c\in\mathbb Z\,.
\end{align}
Following the notation of~(\ref{e:IsigmarhoDefdual}) we have
\begin{equation}
\vec
    J_1 (a,b,c;\eta)=\begin{pmatrix}
{\displaystyle  J_{(Id,\emptyset)}(a,b,c;\eta)} \cr
{\displaystyle J_{(\emptyset,Id)}(a,b,c;\eta)}
\end{pmatrix}=
\begin{pmatrix}\label{e:J1def}
{\displaystyle \int_{-\infty}^0 
  (-z)^{a} (1-z)^{b} (\eta-z)^{c}dz}\cr 
{\displaystyle \int_\eta^1 
  z^{ a} (1-z)^{ b} (z- \eta)^{ c}dz}
\end{pmatrix}\,.
\end{equation}
The conformal block decomposition gives 
\begin{equation}\label{e:holresultJ}
  \mathcal G_1\left({ a\,  b\, c\atop {\tilde a}\,
      {\tilde b}\, {\tilde c}}\Bigg|\eta\right) =
   \frac 1\pi\, \vec  J_1(\tilde a,\tilde b,\tilde c;\bar \eta)^T
\, \hat G_1(a,b,c)\,
\vec J_1 (a,b,c;\eta),
\end{equation}
with 
\begin{equation}\label{e:G1resultJ}
 \hat G_1(a,b,c)=
{-1\over\sin(\pi(b+c))} \begin{pmatrix}
 \sin(\pi(a+b+c))  \sin(\pi a) &0\cr
  0  &\sin(\pi b)\sin(\pi c)
 \end{pmatrix}\,.
\end{equation}

%----------------------------------------------------------------------
\subsection{Closed string amplitudes}

Setting $\eta=1$ in the holomorphic
factorisation leads to a fairly simple expression, because at $\eta=1$
only the first $N!$ rows $\vec \jmath_N$ of the vector of
Aomoto-Gel'fand hypergeometric functions $\vec J_N$
in~\eqref{e:IsigmarhoDefdual} are non-vanishing:\footnote{The lower $N\times N!$ components vanish whenever the integral is absolutely convergent. This leads to~\eqref{e:Jlimit} by analytic
continuation on the parameters $\pmb b$, $\pmb c$ and $\pmb d$.}
\begin{equation}\label{e:Jlimit}
   \vec J_N(\pmb a,\pmb b,\pmb c;\pmb d;1) =
    \begin{pmatrix}
    \vec \jmath_N(\pmb a,\pmb b,\pmb c;\pmb d)   \cr 0
    \end{pmatrix},
  \end{equation}
  where
\begin{equation}\label{e:IsigmarhoDefdualz1}
\vec \jmath_{(\rho
  ,\sigma)}(\pmb a,\pmb b,\pmb c;\pmb
                d)=\int_{\tilde\delta_{(\rho,\sigma)}}
 \prod_{1\leq m<n\leq N}
(z_{\rho(m)}-z_{\rho(n)})^{d_{\rho( m) \rho( n)}}
\prod_{m=1}^{N} (-z_{\rho(m)})^{a_{\rho(m)}} (1-z_{\rho(m)})^{b_{\rho(m)}+{c_{\rho(m)}}}\prod_{i=1}^{N} dz_i\,,
\end{equation}
integrated over the domain
\begin{equation}
  \label{e:DeltaDualDefz1}
  \tilde\delta_{(\rho,\sigma)}:= \{z_{\rho(N)}\leq \cdots\leq
  z_{\rho(1)} \leq 0 \}\,.
\end{equation}
These integrals can be identified with the open string building blocks $A^\rho_{N+3}(\pmb s,\pmb n)$ of~\eqref{Abuildblock}.

Using the parameter identification~(\ref{e:paramA}) and the holomorphic factorization, we find
\begin{equation}\label{e:MnResult}
   M_{N+3}(\pmb s,\pmb n,\pmb {\tilde n})= \mathcal G_{N}\left({\pmb a\, \pmb b\, \pmb c\, \pmb d\atop \pmb{\tilde a}\,
      \pmb{\tilde  b}\, \pmb{\tilde c}\, \pmb{\tilde d}}\Bigg|1\right)=\frac 1{\pi^N}\, \vec \jmath_N(\tilde{\pmb
   a},\tilde{\pmb b},\tilde{\pmb c};\pmb d)^T  \, \hat G_N^{(1)}(\pmb a,\pmb b,\pmb c;\pmb d) \, \vec \jmath_N(\pmb
   a,\pmb b,\pmb c;\pmb d),
\end{equation}
where $\hat G_N^{(1)}(\pmb a,\pmb b,\pmb c;\pmb d)$ is the upper-left $N\times N$ block matrix contained in $\hat G_N(\pmb a,\pmb b,\pmb c;\pmb d)$.

 This approach leads to an expression of the closed string building blocks in terms of the size $N!$ vector $\vec \jmath_N(\pmb
   a,\pmb b,\pmb c;\pmb d) $ of open string building blocks and the matrix $\hat G^{(1)}_N$, which is a non-local
   version of the KLT relations given
   in~\cite{Mizera:2016jhj,Mizera:2017cqs}. One finds back the usual KLT relations~\cite{Kawai:1985xq} using the
   monodromy relations~\cite{BjerrumBohr:2009rd,Stieberger:2009hq} between open string ordered integrals.

Although the end result for the closed string building block only
depends on $\hat G_N^{(1)}$, it is necessary to have the full matrix
$\hat G_N$ to get rid of the monodromies when $\eta$ varies.
   
%--------------------------------------------------------------------------
\subsubsection{The four-point case}
\label{sec:svM4}

We illustrate the previous results on the $N=1$ case corresponding to
the four-point amplitude

  \begin{equation}
      M_4(\pmb s,\pmb n,\pmb {\tilde n})=\int_{\mathbb{P}^1_{\mathbb{C}}}
     | z|^{-2\alpha' s_{12}} z^{n_{12}}\bar z^{\tilde n_{12}}
     |1-z|^{-2\alpha' s_{13}}
     (1-z)^{n_{13}}
     (1-\bar z)^{\tilde n_{13}}
 \,d^2z\,.
    \end{equation}
We can write
\begin{equation}
      M_4(\pmb s,\pmb n,\pmb {\tilde n})=\mathcal G_1\left({a\, b\,
        c\atop \tilde a\,\tilde b\,\tilde c}\Bigg| 1\right),
\end{equation}
with the identification of the parameters 
$  a=-\alpha' s_{12}+n_{12}$, $\tilde a=-\alpha' s_{12}+\tilde n_{12}$,
$b+c=-\alpha' s_{13}+n_{13}$ and $\tilde b+\tilde c=-\alpha' s_{13}+\tilde
n_{13}$. The second component of $\vec J_1(a,b,c;\eta)$ in~\eqref{e:J1def}
vanishes as $\eta\to1$ in the convergence region, and therefore everywhere by analytic continuation.

With such identifications, $j_1(a,b,c)$ is nothing but the integral
\begin{equation}\label{j1A4}
A(1234;\pmb s,\pmb n):=\int_{-\infty}^0(-z)^{ -\alpha' s_{12}+ n_{12}}(1-z)^{-\alpha'  s_{13}+ n_{13}}\,dz \,,
\end{equation}
where $(1234)$ refers to the permutation of the points $z,0,1,+\infty$ on the real line which gives the ordered integral~(\ref{j1A4}). The holomorphic decomposition in~\eqref{e:holresultJ} evaluated
at $\eta=1$ then gives (using the momentum conservation condition $s_{12}+s_{13}+s_{23}=0$)
\begin{equation}\label{e:Gholo4pt}
    M_4(\pmb s,\pmb n,\pmb {\tilde n})= -\frac 1\pi \,{\sin(\alpha' \pi s_{12})
    \sin(\alpha' \pi s_{23})\over  \sin(\alpha' \pi s_{13})} A(1234;\pmb s,\pmb n)A(1234;\pmb s,\pmb {\tilde n}) \,.
\end{equation}
The KLT relation with the inverse momentum kernel
arises from this formalism. The ordered integral $A(1234;\pmb s,\pmb n)$ is just one of the possible four-point open string ordered integrals, but by standard monodromy relations~\cite{BjerrumBohr:2009rd,Stieberger:2009hq} any other four-point ordered integral must be related to $A(1234;\pmb s,\pmb n)$. For instance, if we consider the permutation $(2134)$ corresponding to the integral
\begin{equation}
    A(2134;\pmb s,\pmb n)   :=\int_0^1
                                                     z^{ -\alpha' s_{12}+ n_{12}}
                                       (1-z)^{ -\alpha' s_{13}+ n_{13}}\,dz \,,
\end{equation}
we have the monodromy relation
\begin{equation}
\sin(\alpha'\pi s_{23}) A(1234;\pmb s,\pmb n)  =\sin(\alpha'\pi
s_{13})  A(2134;\pmb s,\pmb n).
\end{equation}
Plugging this relation in~\eqref{e:Gholo4pt} we recover\footnote{The sign difference is given by a different notation for the Mandelstam variables.} the KLT expression
for the closed string four-point amplitudes given  in~\cite[eq.~(3.11)]{Kawai:1985xq}:
\begin{align}
      M_4(\pmb s,\pmb n,\pmb {\tilde n})&=  -\frac 1\pi\sin(\alpha'\pi s_{12})
                                        A(1234;\pmb s,\pmb n)  A(2134;\pmb s,\pmb {\tilde n}) \cr
                                        &= -\frac 1\pi\sin(\alpha'\pi s_{12})
  A(2134;\pmb s,\pmb n)   A(1234;\pmb s,\pmb {\tilde n}) \,.
\end{align}

This illustrates how the momentum kernel $\mathcal S_N$ from~\cite{BjerrumBohr:2010hn} arises as the product of
the holomorphic factorisation matrix $\hat G_N^{(1)}$ times the change
of basis matrix between the ordered open string integrals:
\begin{equation}
  \mathcal S_1(\alpha' s_{12})= -\sin(\alpha'\pi s_{12})  =  -{\sin(\alpha'\pi s_{12})
    \sin(\alpha'\pi s_{23})\over  \sin(\alpha'\pi s_{13})}\times {\sin(\alpha'\pi s_{13})\over  \sin(\alpha'\pi s_{23})}\,.
\end{equation}
The $\Gamma$-function representation of the (beta-) integrals defining $A(1234;\pmb s,\pmb n)$ and $A(2134;\pmb s,\pmb n)$ is
\begin{align}
A(1234;\pmb s,\pmb{\tilde n})&={\Gamma(1-\alpha'  s_{12}+\tilde n_{12})\Gamma(-1-\alpha' s_{23}-\tilde n_{12}-\tilde
n_{13})\over\Gamma(\alpha'  s_{13}-\tilde n_{13})}\cr
 A(2134;\pmb s,\pmb n)   &={\Gamma(1-\alpha' 
                          s_{12}+n_{12})\Gamma(1-\alpha'  s_{13}+n_{13})\over\Gamma(2+\alpha' s_{23}+n_{12}+n_{13})}\,.          
\end{align}

Using the momentum conservation condition $s_{12}+s_{13}+s_{23}=0$ and Euler's reflection formula $\Gamma(x)\Gamma(1-x)=\tfrac \pi{\sin(\pi x)}$, the four-point partial amplitude reads
\begin{equation}
  M_4(\pmb s,\pmb n,\pmb {\tilde n})= -(-1)^{\tilde
    n_{13}}
  {\Gamma(1 -\alpha'  s_{12} + n_{12})\over \Gamma(\alpha'  s_{12} - \tilde n_{12})}
   {\Gamma(1 -\alpha'  s_{13} + n_{13})\over \Gamma(\alpha' 
          s_{13} - \tilde n_{13})}{\Gamma(-1
          -\alpha' s_{23} - \tilde n_{12} - \tilde n_{13})\over \Gamma(2 +\alpha' 
          s_{23} + n_{12} + n_{13})} \,.
\end{equation}
Using repeatedly that
$\Gamma(1+x)=x\Gamma(x)$, one can find a rational function $Q(\alpha's_{12},\alpha's_{13},\alpha's_{23})$  with integer coefficients (depending on the integer parameters $n_{12}, n_{13}, \tilde n_{12}, \tilde n_{13}$) such that
\begin{equation}\label{e:M4Gammaratio}
  M_4(\pmb s,\pmb n,\pmb {\tilde n})= Q(\alpha' s_{12},\alpha' s_{13},\alpha' s_{23})
  {\Gamma(1 -\alpha'  s_{12} )\over \Gamma(1+\alpha'  s_{12})}
   {\Gamma(1 -\alpha'  s_{13})\over \Gamma(1+\alpha' 
          s_{13})}{\Gamma(1
         -\alpha' s_{23})\over \Gamma(1 +\alpha' 
          s_{23})} \,.
\end{equation}
Euler's formula
\begin{equation}
  \Gamma(1+x)= e^{-\gamma x} \exp\left(\sum_{m\geq2} {\zeta
      (m)\over m} (-x)^m\right),
\end{equation}
where $\gamma$ is the Euler-Mascheroni constant, implies that
\begin{equation}
  {\Gamma(1+x)\over \Gamma(1-x)}=
  e^{-2\gamma x}
  \exp\left(-2\sum_{m=1}^\infty {\zeta(2m+1)\over 2m+1} x^{2m+1}\right)  \,.
\end{equation}
Therefore, by momentum conservation, the four-point partial amplitude takes the form
\begin{equation}\label{e:M4exp}
    M_4(\pmb s,\pmb n,\pmb {\tilde n})=Q(\alpha' s_{12},\alpha'
    s_{13},\alpha' s_{23})
\exp\left(2\sum_{m=1}^\infty {\zeta(2m+1)\over 2m+1} \Big(( \alpha' s_{12})^{2m+1}+( \alpha' s_{13} )^{2m+1}+( \alpha' s_{23} )^{2m+1}\Big)\right) \,.
\end{equation}
The argument of the exponential factor only involves odd Riemann
zeta values. As we will see in the next section, the fact that even zeta values do not appear is part of a more general pattern which seems to constitute a fundamental feature of closed string theories.

\section{Small $\alpha'$ expansion of tree-level building blocks}
\label{sec:historic}

The disc amplitude building blocks  with $N+3$ external states
on the boundary~\eqref{e:Acolor} can be reduced (rescaling~$\alpha'$ by a factor of~$4$) to the finite set of
integrals~\cite{Schlotterer:2018zce}

\begin{equation}\label{e:Zdef}
Z_{\rho,\sigma}^{(N)}(\alpha'\pmb s):= 
\int_{0\leq x_{\sigma(1)}\leq\cdots\leq x_{\sigma(N)}\leq
  1}\,\frac{\prod_{1\leq i<j\leq N+2}|x_i-x_j|^{- \alpha's_{ij}}}{x_{\rho(1)}(1-x_{\rho(N)})\prod_{i=2}^N(x_{\rho(i)}-x_{\rho(i-1)})}\,\prod_{i=1}^N
dx_i\,,
\end{equation}
with  $\rho, \sigma \in \mathfrak{S}_N$ permutations of $N$
letters, $x_{N+1}:=0$, $x_{N+2}:=1$.
Similarly, the closed string amplitude building blocks in~\eqref{e:partamplitudes} can be expanded on the
finite set of integrals~\cite{Stieberger:2014hba,Mafra:2011nv,Broedel:2013aza}
\begin{equation}\label{e:Jdef}
J_{\rho,\sigma}^{(N)}(\alpha'\pmb s):= 
\int_{(\mathbb{P}_\mathbb{C}^1)^N}\frac{\prod_{1\leq i<j\leq
    N+2}|z_i-z_j|^{-2\alpha's_{ij} }}{z_{\rho(1)}\overline{z}_{\sigma(1)}(1-z_{\rho(N)})(1-\overline{z}_{\sigma(N)})\prod_{i=2}^N(z_{\rho(i)}-z_{\rho(i-1)})(\overline{z}_{\sigma(i)}-\overline{z}_{\sigma(i-1)})}\prod_{i=1}^N
d^2z_i,
\end{equation}
where $d^2z_i:=idzd\overline{z}/2\pi$, $z_{N+1}:=0$, $z_{N+2}:=1$.

It was shown in~\cite{Broedel:2013aza} that
$Z_{\rho,\sigma}^{(N)}(\alpha'\pmb s)$ have a Laurent
series expansion as $\alpha'$ tends to zero, with coefficients belonging to the ring $\mathcal{Z}$ of rational linear combination of \emph{multiple zeta values}, the real numbers defined for $k_1,\ldots ,k_r\in\mathbb{N}$, $k_r\geq 2$, by the absolutely convergent series
\begin{equation}\label{MZV}
\zeta(k_1,\ldots ,k_r):=\sum_{0<n_1<\cdots <n_r}\frac{1}{n_1^{k_1}\cdots n_r^{k_r}}.
\end{equation}

The functions $J_{\rho,\sigma}^{(N)}(\alpha's_{ij})$ also have a Laurent
series expansion as $\alpha'$  tends to zero, and extensive computations led to conjecture
in~\cite{Stieberger:2013wea} that the coefficients belong to a subring $\mathcal{Z}^{\rm sv}$ of multiple zeta values, called the ring of \emph{single-valued multiple zeta values}, which we will introduce in section~\ref{Sec:svmpls} below. This conjecture was recently proven, independently and in three different ways, in the papers~\cite{Schlotterer:2018zce, Brown:2018omk} and in our paper~\cite{Vanhove:2018elu}. 

The methods of~\cite{Schlotterer:2018zce, Brown:2018omk} lead to
proving a stronger statement, which relates
$J_{\rho,\sigma}^{(N)}(\alpha' s_{ij})$ to
$Z_{\rho,\sigma}^{(N)}(\alpha' s_{ij})$ via
the \emph{single-valued projection}, a surjective map from the
(motivic) ring of multiple zeta values to the (motivic) ring of
single-valued multiple zeta values. Both of these proofs can be used
to explicitly compute the $\alpha'$-expansion of
$J_{\rho,\sigma}^{(N)}(\alpha' s_{ij})$ provided that we know the
$\alpha'$-expansion of $Z_{\rho,\sigma}^{(N)}(\alpha's_{ij})$.

Our proof, instead, does not involve open string integrals, and can
be used to algorithmically compute the coefficients of
$J_{\rho,\sigma}^{(N)}(\alpha's_{ij})$ directly from their
definition. In section~\ref{Sectionk=1} we give a detailed explanation
of our approach in the simplest case where $N=1$. An advantage of our
method is that it allows to compute a broad class of multiple
complex integrals in terms of single-valued multiple zeta values. For
instance, we will see in section~\ref{sec:two-points-function} that one can use the same techniques to prove that certain asymptotic limits of genus-one closed string amplitudes involve single-valued multiple zeta values.

\subsection{Single-valued multiple zeta values}\label{Sec:svmpls}

We consider the alphabet $X=\{x_0,x_1\}$ formed by two formal non-commutative letters~$x_0$ and~$x_1$, and we denote by~$X^*$ the set of all possible words $w=x_{i_1}\cdots x_{i_n}$ in this alphabet. We fix a simply connected complex domain~$U$ obtained by removing from~$\mathbb{C}$ two non-intersecting half-lines (the \emph{branch cuts}) connecting~$\infty$ with~$0$ and~$1$, respectively.

To every word $w\in X^*$ we associate a \emph{multiple polylogarithm} (in one variable) $L_w(z)$, the holomorphic function of $z\in U$ defined recursively by setting $L_{x_0^r}(z):=(\log(z))^r/r!$ (the principal branch on~$U$) for a string~$x_0^r$ of~$r$ consecutives~$x_0$'s and by setting for any other word $w=x_{i_1}\cdots x_{i_n}\in X^*$
\begin{equation}
L_w(z):=\int_{[0,z]}\frac{dz'}{z'-i_1}\, L_{x_{i_2}\cdots x_{i_n}}(z'),
\end{equation}
where $[0,z]$ is any path in~$U$. These functions are well-defined, because they are invariant under deformations of the path  inside the same homotopy class, and thus depend only on the endpoint~$z$, because the fundamental group of~$U$ is trivial. They can be extended to multi-valued functions on $\mathbb{C}\setminus\{0,1\}$, whose values depend on $z$ and on the homotopy class of the path $[0,z]$ inside $\mathbb{C}\setminus\{0,1\}$.

For all~$w$ there exist an integer $K_0\geq 0$ and holomorphic functions $f_0,\ldots ,f_{K_0}$ vanishing at~$0$ such that, if $z\in U$ is close enough to~$0$, then
\begin{equation}\label{asympt0}
L_w(z)=\sum_{k=0}^{K_0}\frac{\big(\log(z)\big)^k}{k!}f_{k}(z).
\end{equation}
Moreover, for all $w$ there exist an integer $K_1\geq 0$ and holomorphic functions $g_0,\ldots ,g_{K_1}$ whose values at~$1$ belong to the ring~$\mathcal{Z}$ of multiple zeta values defined by~(\ref{MZV}) such that, if $z\in U$ is close enough to~$1$, then
\begin{equation}
L_w(z)=\sum_{k=0}^{K_1}\frac{\big(\log(1-z)\big)^k}{k!}g_{k}(z).
\end{equation}
We define the \emph{regularised special values} $L_w(1):=g_0(1)\in\mathcal{Z}$ (which coincide with the limit at $z=1$ if the word $w$ starts with $x_0$). In particular, $\zeta(k_1,\ldots ,k_r)=(-1)^rL_{x_0^{k_r-1}x_1\cdots x_0^{k_1-1}x_1}(1)$.

The monodromies of multiple polylogarithms along paths in $\mathbb{C}\setminus\{0,1\}$ crossing the branch cuts are best described in terms of the generating series
\begin{equation}
L_{\{x_0,x_1\}}(z):=\sum_{w\in X^*}L_w(z)\cdot w.
\end{equation}
This is, up to constants, the unique holomorphic
$\mathbb{C}\langle\langle X \rangle\rangle$-valued solution in $U$ of
the \emph{Knizhnik-Zamolodchikov (KZ) equation}
\begin{equation}
\frac{\partial}{\partial z} F(z)=\bigg(\frac{x_0}{z}+\frac{x_1}{z-1}\bigg) F(z)
\end{equation}
such that $F(z)=\exp(x_0\log(z))\,f(z)$ as $z\rightarrow 0$, with $f$ holomorphic $\mathbb{C}\langle\langle X \rangle\rangle $-valued such that $f(0)=1$. The regularised special value $Z_{\{x_0,x_1\}}:=L_{\{x_0,x_1\}}(1)$, i.e. the generating series of (regularised) multiple zeta values, is known as the Drinfel'd associator. With this notation, the action of the monodromy operators $M_0$ and $M_1$ (defined by analytic continuation along loops winding once counterclockwise around~$0$ and~$1$, respectively) is given by
\begin{equation}
M_0\,L_{\{x_0,x_1\}}(z)=L_{\{x_0,x_1\}}(z)\,e^{2\pi ix_0}, \;\;\;\;\;\;\;\;\; M_1\,L_{\{x_0,x_1\}}(z)=L_{\{x_0,x_1\}}(z)\,Z_{\{x_0,x_1\}}^{-1}\,e^{2\pi ix_1}\,Z_{\{x_0,x_1\}}.
\end{equation}

Looking at the monodromy around $z=0$, a first step towards constructing single-valued analogues of multiple polylogarithms is to define on $X^*$ the operator $\widetilde{x_{i_1}\cdots x_{i_n}}:=x_{i_n}\cdots x_{i_1}$, extend it by linearity to $\mathbb{C}\langle\langle X \rangle\rangle $ and 
consider the product
$L_{\{x_0,x_1\}}(z)\widetilde{\overline{L_{\{x_0,x_1\}}(z)}}$. This new function
is now single-valued around $z=0$, but it is still multi-valued around
$z=1$. It turns out~\cite{BrownSVMPL}  that one can ``deform'' the letter $x_1$ to a new letter $y(x_0,x_1)=x_1+\ldots\in\mathcal{Z}\langle\langle X \rangle\rangle$ so that the new function
\begin{equation}\label{DefSVMPLS}
\mathcal{L}_{\{x_0,x_1\}}(z):=L_{\{x_0,x_1\}}(z)\,\widetilde{\overline{L_{\{x_0,y(x_0,x_1)\}}(z)}}
\end{equation}
is single-valued (but no-longer holomorphic) on the punctured complex plane $\mathbb{C}\setminus\{0,1\}$. It can be seen as a generating function
\begin{equation}
\mathcal{L}_{\{x_0,x_1\}}(z)=:\sum_{w\in X^*}\mathcal{L}_w(z)\cdot w
\end{equation}
of single-valued smooth functions $\mathcal{L}_w(z)$, which are called \emph{single-valued multiple polylogarithms}, whose simplest examples $\mathcal{L}_{x_0^{r}}(z)=(\log|z|^2)^r/r!$ are indeed ``single-valued versions'' of the corresponding multiple polylogarithms $L_{x_0^{r}}(z)=(\log(z))^r/r!$. Remarkably, also the function $\mathcal{L}_{\{x_0,x_1\}}(z)$ is a solution to the KZ equation, and the map $L_w(z)\rightarrow \mathcal{L}_w(z)$ respects shuffle product identities.

\emph{Single-valued multiple zeta values} are defined as the (regularised) special values $\mathcal{L}_w(1)$ obtained by setting $z=1$ in the equation~(\ref{DefSVMPLS}). Because $y(x_0,x_1)\in\mathcal{Z}\langle\langle X \rangle\rangle$, they also belong to $\mathcal{Z}$, but in fact they form a (conjecturally) much smaller subring $\mathcal{Z}^{\rm sv}$. Assuming standard conjectures, the map $L_w(z)\rightarrow \mathcal{L}_w(z)$ specializes at $z=1$ to a surjective map of rings $\mathcal{Z}\rightarrow \mathcal{Z}^{\rm sv}$, which we call the ``single-valued projection''. It therefore seems natural to define $\zeta^{\rm sv}(k_1,\ldots ,k_r):=(-1)^r\mathcal{L}_{x_0^{k_r-1}x_1\cdots x_0^{k_1-1}x_1}(1)$, and one finds for instance $\zeta^{\rm sv}(2k)=0$ and $\zeta^{\rm sv}(2k+1)=2\,\zeta(2k+1)$.

\subsection{The $\alpha'$-expansion of the four-point amplitude from single-valued integration}\label{Sectionk=1}

We illustrate the approach of~\cite{Vanhove:2018elu} to prove that the coefficients of tree-level closed string building blocks are single-valued multiple zeta values in the case of the four-point integral
\begin{equation}\label{4closedStrings}
V(s,t):=\int_{\mathbb{P}^1_{\mathbb{C}}}|z|^{2s-2}|1-z|^{2t-2}d^2z.
\end{equation}
This function coincides with the Virasoro-Shapiro
amplitude~(\ref{e:VirasoroShapiro}) upon setting $s:=-\alpha' s_{12}
+1$ and $t:=-\alpha's_{13} +1$ and coincides with $J^{(1)}_{\rm
  Id,Id}$ from eq.~(\ref{e:Jdef}) upon setting $s:=-\alpha' s_{12}
$ and $t:=-\alpha's_{13}$. The integral defining $V(s,t)$ is  absolutely
convergent for  $s,t\in\mathbb{C}$ such that $\textup{Re}(s)>0$,
$\textup{Re}(t)>0$ and $\textup{Re}(s+t)<1$.

The integrand is entire in $s$ and $t$ as soon as $z$ belongs to the domain $D_\varepsilon$ obtained from $\mathbb{P}^1_{\mathbb{C}}$ by cutting out small discs of radius $\varepsilon$ around $z=0,1,\infty$, where it can therefore be expanded as a power series at $(s,t)=(0,0)$. Hence we can write
\begin{equation}
V(s,t)=\lim_{\varepsilon\rightarrow 0}\int_{D_\varepsilon}\sum_{p,q\geq 0}s^pt^q\,\frac{\mathcal{L}_{x_0^p}(z)\mathcal{L}_{x_1^q}(z)}{|z|^2|1-z|^2}d^2z,
\end{equation} 
where we recall that $\mathcal{L}_{x_0^p}(z)=(\log|z|^2)^p/p!$ and
$\mathcal{L}_{x_1^q}(z)=(\log|1-z|^2)^q/q!$. Exchanging integration
with summation and rearraging the summation\footnote{\label{note}We refer
  to~\cite{Vanhove:2018elu} for a justification of these steps and details.}, one finds
\begin{equation}\label{LimitBeta}
V(s,t)=\lim_{\varepsilon\rightarrow 0}\Bigg(\sum_{p,q\geq 1}s^pt^q
\int_{D_\varepsilon}\frac{\mathcal{L}_{x_0^p}(z)\mathcal{L}_{x_1^q}(z)}{|z|^2|1-z|^2}d^2z
+\sum_{\substack{p,q\geq 0\\p\cdot q=0}}s^pt^q\int_{D_\varepsilon}\frac{\mathcal{L}_{x_0^p}(z)\mathcal{L}_{x_1^q}(z)}{|z|^2|1-z|^2}d^2z\Bigg)\,.
\end{equation}

We focus now on the first term. Because the map $L_w(z)\rightarrow \mathcal{L}_w(z)$ respects shuffle product identities and the KZ equation, we get
\begin{equation}
\frac{\mathcal{L}_{x_0^p}(z)\mathcal{L}_{x_1^q}(z)}{|z|^2|1-z|^2}=\frac{\partial}{\partial z}
\Bigg(\sum_{w=x_0^p\shuffle x_1^q}\frac{\mathcal{L}_{x_0w}(z)-\mathcal{L}_{x_1w}(z)}{\overline{z}(1-\overline{z})}\Bigg).
\end{equation}
Using this fact, the Stokes theorem and the known asymptotic behaviour of multiple polylogarithms at $0,1,\infty$, and denoting by $B_z(\varepsilon)$ a disc centered at $z$ of radius $\varepsilon$, one finds that for any $p,q\geq 1$
\begin{align}
\int_{D_{\varepsilon}}\frac{\mathcal{L}_{x_0^p}(z)\mathcal{L}_{x_1^q}(z)}{|z|^2|1-z|^2}d^2z&\,=\,-\frac{1}{2\pi i}\sum_{z=0,1,\infty}\bigg(\int_{\overline{\partial} B_z(\varepsilon)}\frac{d\overline{z}}{\overline{z}(1-\overline{z})}\sum_{w=x_0^p\shuffle x_1^q}\big(\mathcal{L}_{x_0w}(z)-\mathcal{L}_{x_1w}(z)\big)\bigg) \notag\\
&\,=\,\sum_{w=x_0^p\shuffle x_1^q}\big(\mathcal{L}_{x_0w}(1)-\mathcal{L}_{x_1w}(1)\big) \,+\, O(\varepsilon).
\end{align}
This method to compute integrals of single-valued multiple polylogarithms is sometimes referred to as \emph{single-valued integration} and was introduced by Schnetz in~\cite{Schnetz:2013hqa} (see also~\cite{DelDuca:2016lad} for an application to multi-Regge kinematics). By a more careful analysis of the dependence on $\varepsilon$ one concludes that
\begin{equation}
\sum_{p,q\geq 1}s^pt^q
\int_{D_\varepsilon}\frac{\mathcal{L}_{x_0^p}(z)\mathcal{L}_{x_1^q}(z)}{|z|^2|1-z|^2}d^2z\,=\,\sum_{p,q\geq 1}\bigg(\sum_{w=x_0^p\shuffle x_1^q}\mathcal{L}_{x_0w}(1)-\mathcal{L}_{x_1w}(1)\bigg)\,s^pt^q \,+\, O(\varepsilon).
\end{equation}

The second term of~(\ref{LimitBeta}) is treated similarly. We need to separate the contributions given by the Stokes theorem applied on the boundaries of the three discs $B_0(\varepsilon), B_1(\varepsilon), B_\infty(\varepsilon)$. It turns out${}^{\cref{note}}$ that the contribution from $B_0(\varepsilon)$ is
\begin{equation}
  -\sum_{p\geq0}
   \left(\frac{(\log\varepsilon^2)^{p+1}}{(p+1)!}\right)\,s^p\,+\,O(\varepsilon)\, =\,\frac{1}{s}(1-\varepsilon^{2s})\,+\,O(\varepsilon),
\end{equation}
the contribution from $B_\infty(\varepsilon)$ is $O(\varepsilon)$ and the contribution from $B_1(\varepsilon)$ is
\begin{align}
  &\sum_{q\geq 1}\mathcal{L}_{x_0x_1^q}(1)\,t^q-\sum_{q\geq0}
   \left(\frac{(\log\varepsilon^2)^{q+1}}{(q+1)!}\right)t^q-\sum_{p\geq 1}\mathcal{L}_{x_1x_0^p}(1)\,s^p + O(\varepsilon) \\
   &=\,\sum_{q\geq 1}\mathcal{L}_{x_0x_1^q}(1)\,t^q+\frac{1}{t}(1-\varepsilon^{2t})-\sum_{p\geq 1}\mathcal{L}_{x_1x_0^p}(1)\,s^p + O(\varepsilon).\notag
\end{align}
Taking the limit $\varepsilon\rightarrow 0$ and noting that $\mathcal{L}_{x_0x_1^n}(1)=-\mathcal{L}_{x_1x_0^n}(1)=\zeta^{\rm sv}(n+1)$ for $n\geq 1$, we obtain 
\begin{equation}
V(s,t)=\frac{s+t}{st}+2\sum_{n\geq 1}\zeta(2n+1)(s^{2n}+t^{2n})+\sum_{p,q\geq 1}\bigg(\sum_{w=x_0^p\shuffle x_1^q}\mathcal{L}_{x_0w}(1)-\mathcal{L}_{x_1w}(1)\bigg)\,s^pt^q.
\end{equation}
We deduce that $V(s,t)$ can be analytically continued to define a meromorphic function in a neighborhood of $(s,t)=(0,0)$, and the coefficients of its Laurent series expansion at that point belong to the ring $\mathcal{Z}^{\rm sv}$ of single-valued multiple zeta values. 

Of course, this result was already clear from eq.~(\ref{e:M4exp}), which implies the stronger fact that the coefficients of $V(s,t)$ are polynomials in odd zeta values. The advantage of the computation above is that it can be generalized to higher-point amplitudes. To do this, the idea is to apply the single-valued
integration recursively, but it is necessary to overcome two difficulties. The first is that the singularity divisors coming from the integrand are not normal crossing, and this can be solved with standard geometric or analytic methods~\cite{Brown:2018omk, Vanhove:2018elu}. The second is that it is necessary to work with a more
general set of functions called \emph{single-valued hyperlogarithms}, defined by Brown in~\cite{BrownNotes}, which are analogues of single-valued multiple polylogarithms arising from more general alphabets $X$, whose letters correspond to punctures on the Riemann sphere. The successive
integration of these functions requires to recursively remove the
integration variables from the alphabet. Methods to solve this kind
of problem were developed in~\cite{Vanhove:2018elu} (precisely to prove that the coefficients of closed string tree-level amplitudes are
single-valued multiple zeta values) or, with a
different approach relying on standard conjectures on multiple zeta
values, in~\cite{DelDuca:2016lad}.

\part{One-loop amplitudes}
\label{sec:genus-one}

\section{Closed string building blocks and modular graph functions}

At one-loop, closed oriented string amplitudes are defined on
the complex torus, and can be
decomposed on building blocks as~\cite{DHoker:1988pdl,Green:1987sp,Polchinski:1998rr}
\begin{equation}
  M_{N+1}(\pmb s,\pmb \epsilon)= \sum_r c^{\rm 1-loop}_r(  \pmb s,\pmb
  \epsilon)  M^{\rm 1-loop}_{N+1}(\pmb s,\pmb n^r,\pmb {\tilde n}^r),
\end{equation}
with the one-loop building blocks given by the integrals over the
moduli space of genus-one closed Riemann surfaces\footnote{Bosonic
  strings are not consistent at loop order because of the tachyons in the
  spectrum. For superstring theory amplitudes, at
  higher genus the projection of the supermoduli integral onto the
  bosonic moduli space of genus $g$ curves with marked point cannot
  be done using a naive
  integration of the fermonic variables~\cite{Witten:2012bh,Donagi:2013dua}, and may be achieved
  using Sen's fermionic integration~\cite{Sen:2014pia,Sen:2015hia}. } (the complex tori) with $N+1$ marked points $\mathfrak{M}_{1,N+1}$. This integration domain can be factorized, so that the building blocks take the form
\begin{equation}\label{Mfull}
  M^{\rm 1-loop}_{N+1}(\pmb s,\pmb n,\pmb{\tilde n})= 
  \int_{\mathfrak{M}_{1,1}} {d\tau_1 d\tau_2\over\tau_2^2} \, \hat M^{\rm 1-loop}_{N+1}(\pmb s,\pmb n,\pmb{\tilde n};\tau),
\end{equation}
with the partial (configuration-space) amplitude building blocks
\begin{multline}\label{e:Mpartial}
\hat   M^{\rm 1-loop}_{N+1}(\pmb s,\pmb n,\pmb{\tilde n};\tau)\,=\\ 
 \int_{\mathbb
    T^N} \prod_{i=1}^N {idz_id\overline{z}_i\over
    2\tau_2}\!\!\!\prod_{1\leq i<j\leq N+1} \!\!\!  \!\!\! e^{2\alpha's_{ij} G^{\mathbb T}(z_i,z_j;\tau)}\left(\partial_{z_i} G^{\mathbb
    T}(z_i,z_j;\tau)\right)^{-2n_{ij}} \left(\partial_{\bar z_i} G^{\mathbb
    T}(z_i,z_j;\tau)\right)^{-2\tilde n_{ij}},
\end{multline}
where we fix $z_{N+1}:=0$. The exponents
$n_{ij}$ and $\tilde n_{ij}$ are non-positive integers
arising from the operator product expansion.  The Arakelov scalar Green function  $G^{\mathbb{T}}(z,z';\tau)$ on the complex torus $\mathbb{T}:=\mathbb{C}/(\mathbb{Z}+\tau\mathbb{Z})$ of modulus $\tau=\tau_1+i\tau_2\in\mathbb{H}$ is defined by
\begin{equation}
\label{3b0}
\partial_z\partial_{\bar z} G^{\mathbb{T}}(z,z';\tau) = - \frac{\pi}{2} \delta ^{(2)} (z-z') + { \pi \over 2\tau_2}, \hskip 1in \int_{\mathbb{T}} \, G^{\mathbb{T}}(z,0;\tau)\,dzd\overline{z}=0\, .
\end{equation}
This determines the Green function to be given by
\begin{equation}\label{e:Gclosed}
  G^{\mathbb{T}}(z,z';\tau)= -\frac{1}{2}\log\left|\theta_1(z-z',\tau)  \over
    \eta(\tau)\right|^2+{\pi \Im\textrm{m}(z-z')^2\over \tau_2}\,,
\end{equation}
with the ratio of the odd Jacobi $\theta_1$-function and the Dedekind $\eta$-function given, setting $q= e^{ 2 \pi i \tau}$, by 
\begin{equation}\label{thetaeta}
\frac{\theta_1(z,\tau)}{\eta(\tau)}\,=\,iq^{1\over 12}e^{-i\pi z}(1-e^{2\pi i z})\prod_{n\geq1}(1-q^ne^{2i\pi z})
(1-q^ne^{-2i\pi z})\,.
\end{equation}

We have used the Bern-Kosower representation of one-loop amplitudes
with only first derivatives of the
Green function after using integration by parts in the
integrand~\cite{Bern:1987tw}.
This decomposition is the one-loop equivalent of the
decomposition of tree-level amplitudes in~\eqref{Firsteq} on  the
building blocks~\eqref{e:partamplitudes}.

In this paper we will focus on the partial amplitude building blocks
in~\eqref{e:Mpartial}, rather than the moduli space integrals giving
the full amplitude building blocks in~(\ref{Mfull}). The coefficients
of the small~$\alpha'$ expansion of the partial amplitudes building
blocks are given by \emph{modular graph
  functions}~\cite{DHoker:2015wxz, Green:2008uj} and \emph{modular
  graph forms}~\cite{DG16}, which constitute interesting new classes
of real-analytic modular functions and forms,
respectively. For the
purposes of this paper, it is sufficient to focus on modular graph
functions, which is the topic of the next sections. In particular, in
section~\ref{sec:two-points-function} we present a new application of
the single-valued integration from section~\ref{Sectionk=1} to compute
their asymptotic expansion in the two-point case.

\subsection{Modular graph functions}
\label{sec:modul-graph-funct}

 \begin{figure}[h]
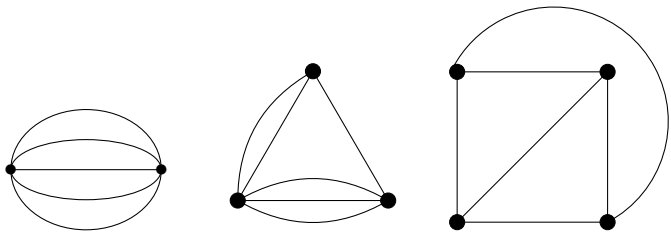

\tikzpicture[scale=2]
\scope[xshift=-5cm,yshift=-0.4cm]
\draw (1,0.0) node{$\bullet$}   ;
\draw (2,0) node{$\bullet$} ;
\draw (1,0) -- (2,0) ;
\draw (1.5,0.0) ellipse (0.5 and 0.2);
\draw (1.5,0.0) ellipse (0.5 and 0.4);
\endscope
\endtikzpicture
\qquad
  \tikzpicture[scale=2]
\scope[xshift=-5cm,yshift=-0.4cm]
\draw (0,0) -- (1,0) ;
\draw (0,0) -- (.5,.86);
\draw (.5,.86) -- (1,0) ;
\draw (0,0) to[bend left] (.5,.86);
\draw (0,0) to[bend left] (1,0);
\draw (0,0) to[bend right] (1,0);
\draw [fill=black]  (0,0)  circle [radius=.05] ;
\draw [fill=black] (1,0)  circle [radius=.05] ;
\draw [fill=black] (.5,.86)  circle [radius=.05] ;
\endscope
\endtikzpicture
\qquad
\tikzpicture[scale=2]
\scope[xshift=-5cm,yshift=-0.4cm]
\draw (0,0) -- (1,0) ;
\draw (0,0) -- (0,1);
\draw (1,1) -- (1,0) ;
\draw (1,1) -- (0,1);
\draw (0,0) -- (1,1) ;
\draw (1,0) arc (-62:152:.76cm);
\draw [fill=black]  (0,0)  circle [radius=.05] ;
\draw [fill=black] (1,0)  circle [radius=.05] ;
\draw [fill=black] (0,1)  circle [radius=.05] ;
\draw [fill=black] (1,1)  circle [radius=.05] ;
\endscope
\endtikzpicture
\caption{Examples of graphs associated to modular graph functions}
\end{figure} 

We consider a graph\footnote{More precisely, we consider an undirected graph with multiple edges connecting pairs of vertices and no self-edges.}~$\Gamma$ drawn on a fundamental domain $\Sigma_\tau\subset\mathbb{C}$ of the torus. Let~$z_i$ with $i=1,\dots,N+1$ be the vertices of the
graph. By translation invariance, we can suppose that $0\in\Sigma_\tau$ and fix $z_{N+1}=0$. We denote by $e_{ij}\geq0$ the number of edges
connecting the points~$i$ and~$j$. The modular graph functions~\cite{DHoker:2015wxz} are defined by the following integrals
associated to the graph~$\Gamma$:

\begin{equation}\label{e:DGammadef}
D_\Gamma(\tau) =\int_{\Sigma_\tau^{N}} \prod_{1\leq i<j\leq N+1}
\big(G^{\mathbb{T}}(z_i,z_j;\tau)\big)^{e_{ij}}\,\prod_{i=1}^N {idz_id\overline{z}_i\over 2\tau_2}.
\end{equation}

The word \emph{modular} refers to the easily checked fact that modular graph functions are invariant under the standard action of $\text{SL}_2(\mathbb{Z})$ on $\tau\in\mathbb{H}$. 
These functions arise naturally in the $\alpha'$-expansion of the partial
amplitude building blocks in~\eqref{e:Mpartial} with $n_{ij}=\tilde n_{ij}=0$. When 
$n_{ij}$ and $\tilde n_{ij}$  are non-vanishing, one gets modular
graph forms~\cite{DG16}, which have non-trivial holomorphic and anti-holomorphic modular weights with respect to the action of $\text{SL}_2(\mathbb{Z})$.

It was proven in~\cite{Zerbini:2018sox} that, if we denote by $E$ the total number of edges of $\Gamma$, the asymptotic expansion as $\tau\rightarrow i\infty$ of modular graph functions takes the form
\begin{equation}\label{asympmodgr}
D_\Gamma (\tau)=\sum_{k=1-E}^E\, \sum_{m,n\geq 0} d_k^{(m,n)}(\Gamma)\, Y^k \,q^m\,\overline{q}^n,
\end{equation}
where $Y:=\pi \tau_2$ and $d_k^{(m,n)}(\Gamma)$ belong to the ring generated by all special values at roots of unity of multiple polylogarithms. A refinement of this theorem was recently announced by E. Panzer~\cite{PanzerTalk}, who proved that $d_k^{(m,n)}(\Gamma)$ belong to the ring~$\mathcal{Z}$ of multiple zeta values. Conjecturally, however, this is still not the optimal result, as the coefficients $d_k^{(m,n)}(\Gamma)$ are expected to belong to the smaller ring~$\mathcal{Z}^{\rm sv}$ of single-valued multiple zeta values~\cite{Zerbini:2015rss,DHoker:2015wxz}. It is presumably straightforward to see that all this applies as well to modular graph forms.

The first Laurent polynomial in this expansion, obtained setting $m=n=0$, is usually called the ``constant term''. Its knowledge is particularly relevant because it can be used to deduce algebraic and differential relations among modular graph functions (and forms)~\cite{DGV1,DG16,DHokerKaidi,GKS1, Gerken:2020aju}, but its computation is extremely hard for graphs with more than two loops or two points~\cite{Zerbini:2015rss,DHoker:2017zhq}. From the geometric viewpoint, it is the main term as $\tau\rightarrow i\infty$, which is the degeneration limit of the complex structure of the torus towards a sphere with two extra marked points. For this reason, it is expected to be related to tree-level amplitudes on a sphere, and this would explain the conjectural relation to single-valued multiple zeta values. We will see in the next section that these expectations can be verified in the simple case of two-point graphs.

\subsection{The banana graphs}
\label{sec:two-points-function}

We look here at the constant term of the asymptotic expansion in the case of two-point modular ``banana graph''
functions
$
D_{n}(\tau)=\hspace{-1.6cm}
\begin{gathered}\tikzpicture[scale=1.7]
\scope[xshift=-5cm,yshift=-0.4cm]
\draw(0.0,0) node{};
\draw (1,0.0) node{$\bullet$}   ;
\draw (2,0) node{$\bullet$} ;
\draw (1,0) -- (2,0) ;
\draw (1.5,0.0) ellipse (0.5 and 0.1);
\draw (1.5,0.0) ellipse (0.5 and 0.2);
\draw (1.5,0.0) ellipse (0.5 and 0.3);
\draw (1.5,0.0) ellipse (0.5 and 0.4);
\endscope
\endtikzpicture
\end{gathered}
$, where~$n$ counts the number of edges, which is one of the simplest families of graphs. It was
shown in~\cite{DHoker:2019xef,Zagier:2019eus}  that 
the coefficients~$d_k^{(0,0)}$ of the constant term can be written in terms of the coefficients of the four-point
closed string amplitude in eq.~\eqref{e:VirasoroShapiro}, which belong to the ring $\mathcal{Z}^{\rm sv}$ as shown in section~\ref{Sectionk=1} and which are actually contained in the smaller subring generated by odd zeta values, as diplayed in eq.~(\ref{e:M4exp}).

We give here an alternative proof that the coefficients $d_{k}^{(0,0)}$ are single-valued multiple zeta values exploiting the
single-valued integration already used
in section~\ref{Sectionk=1}. This proof is inspired by the approach of~\cite{DHoker:2019xef}.

A generating series for the two-point
functions $D_n(\tau)$ is 
\begin{equation}\label{hhhh}
 F(s,\tau)\,:=\,\sum_{n\geq0}{ s^n\over n!} D_n(\tau)
\,=\,\int_{\Sigma_\tau} e^{s G^{\mathbb T}(z,0;\tau)} \,\,{idzd\overline{z}\over 2\tau_2}  
\end{equation}
where $G^{\mathbb T}(z,z';\tau)$ is the genus one closed string Green function defined in~\eqref{e:Gclosed}. The integral on the right hand side coincides with the two-point closed string building block $\hat{M}_{2}^{\text{1--loop}}(\pmb s,\pmb 0,\pmb 0;\tau)$ upon setting $s:=2\alpha's_{12}$.

As in the previous section, we set $Y:=\pi\tau_2$. Moreover, we denote $\alpha:=\Re\textrm{e}(z)$ and $\beta:=\Im\textrm{m}(z)/\tau_2$ so that the standard measure $\tfrac{idzd\overline{z}}{2\tau_2}$ on $\Sigma_{\tau}$ can be written as $d\alpha d\beta$. Finally, we fix the fundamental domain $\alpha\in [0,1]$, $\beta\in [-\frac12,\frac12]$. In this domain, all contributions to the integral coming from the infinite product $\prod_{n\geq 1}\cdots$ in equation~(\ref{thetaeta}) can be ignored for large $Y$, and we find
\begin{equation}\label{firststepF}
 F(s,\tau)=\int_0^1 e^{2 Y s(\beta^2-\beta+\frac16)}
  d\beta+
  e^{sY/3}\int_{-\frac12}^{\frac12}\int_0^1  e^{2Ys\beta^2}e^{-2Ys\beta}
  \left(\left|1-e^{2\pi i \alpha-2Y\beta}\right|^{-2s}-1\right) d\alpha d\beta+O(e^{-2Y})\,.
\end{equation}
The first integral gives the leading terms of the Laurent polynomials (see~\cite[\S D.1]{Green:2008uj}):
\begin{equation}
\int_0^1 e^{2 Y s(\beta^2-\beta+\frac16)}
  d\beta\,= \,\sum_{n\geq0}{ (Ys)^n\over 3^n  n!}
  {}_2F_1\left({1,\, -n\atop \frac32}\Big|\frac32\right)\,,
\end{equation}
where  $ {}_2F_1\left({a\, b\atop c}| z\right)= \sum_{n\geq0}
{\Gamma(a+n)\Gamma(b+n)\Gamma(c)\over
  \Gamma(a)\Gamma(b)\Gamma(c+n)}{z^n\over n!}$ is the Gauss
hypergeometric function.

In the second integral, we make the change of variables $\zeta=\exp(2\pi i\alpha-2Y\beta)$, we Taylor expand the integrand and we exchange integration with summation (by holomorphicity at $s=0$) to obtain
\begin{equation}
{e^{sY/3}\over 4 Y}\sum_{k,n\geq0\atop m\geq1} {(2k+n)!\over k!n!}{(-1)^m\over2^{n+3k}} {s^{k+m+n}\over Y^k} \int_{|q|^{\frac12}\leq |\zeta|\leq |q|^{-\frac12}} {\mathcal L_{x_0^{2k+n}}(\zeta) \mathcal L_{x_1^m}(\zeta)\over |\zeta|^2}\,d^2\zeta\,+\,O(e^{-2Y})\,,
\end{equation}
where we employ the notation $\mathcal L_{x_0^r}(\zeta)=(\log |\zeta|)^r/r!$ and $\mathcal L_{x_1^r}(\zeta)=(\log |1-\zeta|)^r/r!$ from section~\ref{Sec:svmpls} and we denote $d^2\zeta:=id\zeta d\overline{\zeta}/2\pi$.
 
Because the map $L_w(z)\rightarrow \mathcal{L}_w(z)$ respects shuffle product identities and the KZ equation, we get
      \begin{equation}
      {\mathcal
          L_{x_0^{2k+n}}(\zeta) \mathcal L_{x_1^m}(\zeta)\over |\zeta|^2} \,d^2\zeta=
        \sum_{w \in x_0^{2k+n}\shuffle x_1^m}  {\mathcal L_{w}(\zeta) \over|\zeta|^2}\,d^2\zeta
        =d\left(   \sum_{w \in 0^{2k+n}\shuffle
            1^m}   {\mathcal L_{0w}(\zeta) \over\overline{\zeta}} {i\,d\bar\zeta\over2\pi}\right),
      \end{equation}
hence by applying the Stokes theorem we find 
\begin{multline}\label{e:Intphase}
\int_{|q|^{\frac12}\leq |\zeta|\leq |q|^{-\frac12}} {\mathcal
          L_{x_0^{2k+n}}(\zeta) \mathcal L_{x_1^m}(\zeta)\over |\zeta|^2}\,d^2\zeta=\\
=\,\int_0^{2\pi}  \sum_{w \in x_0^{2k+n}\shuffle\,x_1^m} \mathcal L_{x_0w}\left(|q|^{-\frac12} e^{i\theta}\right)\frac{d\theta}{2\pi}\,-\,\int_0^{2\pi}  \sum_{w \in x_0^{2k+n}\shuffle\,x_1^m} \mathcal L_{x_0w}\left(|q|^{\frac12} e^{i\theta}\right)\frac{d\theta}{2\pi}.
\end{multline}

Integrating term-by-term the asymptotic expansion of single-valued multiple polylogarithms at~0, which can be easily deduced from~(\ref{asympt0}), one finds that the second integral on the right hand side is $O(e^{-2Y})$. As for the first integral, since we are now interested in large $Y=\log|q|^{-1/2}$, we need to know the asymptotic behaviour of single-valued multiple polylogarithms at $\infty$. By Lemma~2.17 of~\cite{Schnetz:2013hqa} we have the generating series identity 
\begin{equation}
\mathcal{L}_{\{x_0,x_1\}}(z^{-1})=\mathcal{L}_{\{-x_0-x_1,x_1\}}(z)\mathcal{L}_{\{x_0,-x_0-x_1\}}(1),
\end{equation}
which together with the asymptotic behaviour at~0 implies that there exists an integer $K_{\infty}\geq 0$ (depending on $w$) such that in a neighbourhood of $\infty$ 
\begin{equation}
  \mathcal L_w (z)=\sum_{i,j\geq0}\sum_{k=0}^{K_\infty}
  c_{i,j,k}^{(\infty)}{(\log|z|^2)^{k}  \over z^i \bar z^j}
\end{equation} 
with $ c_{i,j,k}^{(\infty)}\in\mathcal{Z}^{\rm sv}$. Combining this with~\eqref{e:Intphase} implies that there exist an integer $R(k,m,n)\geq 0$ and numbers $Z_r(k,m,n)\in\mathcal{Z}^{\rm sv}$ such that
  \begin{equation}
   \int_{|q|^{\frac12}\leq |\zeta|\leq |q|^{-\frac12}} {\mathcal
          L_{x_0^{2k+n}}(\zeta) \mathcal L_{x_1^m}(\zeta)\over |\zeta|^2}\,d^2\zeta\,=\, \sum_{r=0}^{R(k,m,n)} Z_r(k,m,n) \,(2Y)^r\, +\,O(e^{-2Y}).
\end{equation}

We conclude that for large $Y$ we have 
\begin{align}
F(s,\tau)&=   \sum_{n\geq0}{ (Ys)^n\over 3^n n!}
  {}_2F_1\left({1,\, -n\atop \frac32}\Big|\frac32\right)\notag\\
&+{e^{sY/3}\over 4Y}
        \sum_{k,n\geq0\atop m\geq1} {(2k+n)!\over k!n!}
        {(-1)^m\over2^{n+3k}} {s^{k+m+n}\over Y^k}  \sum_{r=0}^{R(k,m,n)} Z_r(k,m,n) \,(2Y)^r\,+\,O(e^{-2Y})\,.
\end{align}

This proves that the coefficients of the constant term in the asymptotic expansion of the two-point graph functions $D_n(\tau)$ belong to the ring of single-valued multiple zeta values. Even though this proof does not explicitly relate these numbers with the coefficients of a genus-zero amplitude, and does not imply the stronger statement that they belong to the subring generated by odd Riemann zeta values, we believe that it may be of interest because of its potential to generalize to a higher number of points, thanks to the generality of the single-valued integration developed in~\cite{Vanhove:2018elu}.

\section{Open string building blocks and open questions}
\label{sec:holograph}

\begin{figure}[h]
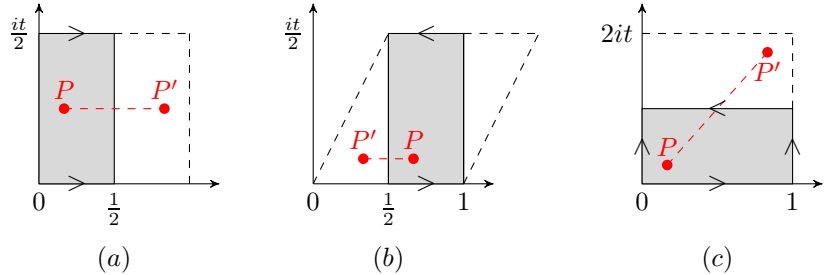
\label{fig:open}
\tikzpicture[scale=2]
\scope[xshift=-5cm,yshift=-0.4cm]
\draw[->] (0,0) to (1.2,0) ;
\draw[->] (0,0) to (0,1.2) ;
\draw[fill=gray!30] (0,0)  rectangle   (1/2,1);
\fill[color=red] (1/6,1/2) circle (1pt);
\draw[color=red] (1/6,1/2) node[above] {$P$};
\fill[color=red] (5/6,1/2) circle (1pt);
\draw[color=red] (5/6,1/2) node[above] {$P'$};
\draw[dashed,color=red] (1/6,1/2) --(5/6,1/2);
\draw (0,0) node[below]{$0$};
\draw (1/2,0) node[below]{$\frac12$};
\draw (0,1) node[left]{${it\over2}$};
\draw (1/4,1) node{$>$};
\draw (1/4,0) node{$>$};
\draw (1/2,-.5) node{$(a)$};
\draw[dashed] (1/2,1) -- (1,1);
\draw[dashed] (1,0) -- (1,1);
\endscope
\endtikzpicture
\qquad
  \tikzpicture[scale=2]
\scope[xshift=-5cm,yshift=-0.4cm]
\draw[->] (0,0) to (1.2,0) ;
\draw[->] (0,0) to (0,1.2) ;
\draw[fill=gray!30] (1/2,0)  rectangle   (1,1);
\draw (1/2,0) to (1/2,1) ;
\draw (1,0) to (1,1) ;
\fill[color=red] (1/3,1/6) circle (1pt);
\draw[color=red] (1/3,1/6) node[above] {$P'$};
\fill[color=red] (2/3,1/6) circle (1pt);
\draw[color=red] (2/3,1/6) node[above] {$P$};
\draw[dashed,color=red] (1/3,1/6) --(2/3,1/6);
\draw (0,0) node[below]{$0$};
\draw (1/2,0) node[below]{$\frac12$};
\draw (1,0) node[below]{$1$};
\draw (3/4,1) node{$<$};
\draw (3/4,0) node{$>$};
\draw[dashed] (0,0) -- (1/2,1);
\draw[dashed] (1,0) -- (3/2,1);
\draw[dashed] (1,1) -- (3/2,1);
\draw (0,1) node[left]{${it\over2}$};
\draw (1/2,-0.5) node {$(b)$};
\endscope
\endtikzpicture
\qquad
\tikzpicture[scale=2]
\scope[xshift=-5cm,yshift=-0.4cm]
\draw[->] (0,0) to (1.2,0) ;
\draw[->] (0,0) to (0,1.2) ;
\draw[fill=gray!30] (0,0)  rectangle   (1,1/2);
\draw (0,0) node[below]{$0$};
\draw (1,0) node[below]{$1$};
\draw (0,1) node[left]{$2it$};
\fill[color=red] (1/6,1/8) circle (1pt);
\draw[color=red] (1/6,1/8) node[above] {$P$};
\fill[color=red] (5/6,7/8) circle (1pt);
\draw[color=red] (5/6,7/8) node[below] {$P'$};
\draw[dashed,color=red] (1/6,1/8) --(5/6,7/8);
\draw[dashed] (0,1) -- (1,1);
\draw[dashed] (1,1/2) -- (1,1);
\draw (1/2,1/2) node{$<$};
\draw (1/2,0) node{$>$};
\draw (0,1/4) node[rotate=90]{$>$};
\draw (1,1/4) node[rotate=90]{$>$};
\draw (1/2,-0.5) node {$(c)$};
\endscope
\endtikzpicture
\caption{Grey area represents the fundamental cell of (a)  the
  annulus $\mathcal{A}$, (b) the M\"obius strip $\mathcal{M}$ and  (c) the Klein bottle $\mathcal{K}$. The points $P$ and $P'$ are identified under the action of the involution on
the covering torus. For the annulus and M\"obius strip the involution is  $\mathcal I_\mathcal{A}(z)=\mathcal I_\mathcal{M}(z)=1-\bar
z$  and for the Klein bottle $\mathcal I_\mathcal{K}(z)=1-\bar z+{\tau_\mathcal{K}\over2}$.}
\end{figure} 

The remaining one-loop topologies are the annulus (open oriented sector), the M\"obius strip (open unoriented sector) and the Klein bottle (closed unoriented sector). The associated amplitudes can be written in terms of building blocks given by integrals on the moduli spaces of annuli, M\"obius strips and Klein bottles~\cite{Mafra:2012kh,Green:2013bza}. Similarly to the torus case, one can factorise the integration domains of the building blocks and obtain partial amplitude building blocks given by integrals over configuration spaces of points on the relevant surface. The form of these integrals strongly depends on the topology of the worldsheet. We start by reviewing the involutions on the covering torus and the Green functions obtained with the method of images, following~\cite{Antoniadis:1996vw}.

\medskip

\noindent{$\bullet$}
The annulus $\mathcal A$ is obtained by the action of the involution
$\mathcal I_{\mathcal{A}}(z)=1-\bar z$ on the doubly covering torus of modulus  $\tau_\mathcal{A}={i t\over2}$ with $t\in\mathbb
R^+$, as represented in
figure~\ref{fig:open}(a). 
The two boundaries of the annulus are at $\Re\textrm{e}(z)=0$ and  $\Re\textrm{e}(z)=\frac12$, and a fundamental cell is given by $[0,\frac12]\times [0,{t\over2}]$.
The method of images applied to the torus propagator in~\eqref{e:Gclosed}, together with well-known properties of $\theta_1$, gives
\begin{equation}\label{e:Gann}
  G^{\mathcal  A}(z,z';\tau_\mathcal{A})=  G^{\mathbb T}(z,z';\tau_\mathcal{A})+ G^{\mathbb T}(
  z, 1-\bar z';\tau_\mathcal{A})\,.
\end{equation}
\noindent{$\bullet$}
The M\"obius strip is obtained by the action of the 
involution $\mathcal I_\mathcal{M}(z)=1-\bar z$ on the doubly covering torus of modulus $\tau_\mathcal{M}=
\frac12+i{t\over2}$ with $t\in\mathbb R^+$, as represented in
figure~\ref{fig:open}(b). The boundary of the M\"obius strip is located at $\Re\textrm{e}(z)=\frac12$ and $\Re\textrm{e}(z)=1$, and a fundamental cell is given by $[\frac12,1]\times [0,{t\over2}]$.
Similarly to the annulus case, the method of images applied to the torus propagator in~\eqref{e:Gclosed}
gives
\begin{equation}\label{e:GMob}
  G^{\mathcal M}(z,z';\tau_\mathcal{M})=   G^{\mathbb T}(z,z';\tau_\mathcal{M})+ G^{\mathbb
    T}(z,1-\bar z';\tau_\mathcal{M})=G^{\mathcal A}(z,z';\tau_\mathcal{M})\,.
\end{equation}

\noindent{$\bullet$}
The Klein bottle $\mathcal K$ is obtained by the action of the involution $\mathcal
I_\mathcal{K}(z)=1-\bar z+{\tau_\mathcal{K}\over2}$  on the doubly covering torus of modulus $\tau_\mathcal{K}=2it$ with $t\in\mathbb R^+$, as represented in
figure~\ref{fig:open}(c). There are no boundary components, and a fundamental cell is given by $[0,1]\times [0,t]$.
The method of images applied to the torus propagator in~\eqref{e:Gclosed}
this time gives
\begin{equation}\label{e:GKlein}
  G^{\mathcal K}(z,z';\tau_\mathcal{K})=G^{\mathbb T}(z,z';\tau_\mathcal{K})+G^{\mathbb
    T}(z, 1-\bar z+{\tau_\mathcal{K}\over2};\tau_\mathcal{K})\,.
\end{equation}

\medskip

These similar expressions for the Green functions suggest that the respective (partial\footnote{As in the torus case, we are mostly interested in the partial amplitudes, given by integrals over configuration spaces.}) amplitudes could be related to each others. It was argued in various cases that, indeed, one can reduce M\"obius strip and Klein bottle partial amplitudes to annulus partial amplitudes~\cite{Green:1984ed, Cappiello:1998er}. As we will discuss in section~\ref{sec:open}, however, very little is known about these amplitudes in all cases which involve closed string states. 

The only case which was systematically studied is that of open string partial amplitude building blocks given by ordered integrals on the boundaries of the annulus or of the M\"obius strip. We report in the next section some recent observation about their relation to the closed string building blocks~(\ref{e:Mpartial}) via the same single-valued projection mentioned at tree-level.

%%%%%%%%%%%%%%%%%%%%%%%%%%%%%%%%%%%%%%%%%%%%%%%%%%%%%%%%%%%%%%%%%
\subsection{Open string partial amplitudes on the annulus}
\label{sec:esv}
%%%%%%%%%%%%%%%%%%%%%%%%%%%%%%%%%%%%%%%%%%%%%%%%%%%%%%%%%%%%%%%%%

One-loop open string partial amplitudes are given by ordered integrals on the boundaries of M\"obius strips and annuli. The annulus case can be divided into the planar case, with string states only on one boundary, and the non-planar case involving both boundaries. The M\"obius strip case can be reduced to the planar annulus case, as explained for instance in~\cite{BMRS}. Moreover, as remarked in~\cite{BMRS, Mafra:2019xms,Broedel:2018izr}, the non-planar annulus case seems to share all the mathematical features of the planar case. For this reason, we only focus on partial amplitude building blocks with all insertions on one boundary of the annulus. We fix these building blocks\footnote{We integrate over the B-cycle of the torus, instead of the A-cycle used in most recent references, to be consistent with our choice of involution given in figure~\ref{fig:open}(a).} to be given by the following iterated integrals over the straight path $[0,\tau]$ with $\tau=\tau_{\mathcal{A}}=it/2\in i\mathbb{R}^+$ (or any other fundamental domain of the annulus boundary $\tau\mathbb{R}/\tau\mathbb{Z}$ depicted in figure~\ref{fig:open}(a)):
\begin{multline}\label{partamplopenone}
\hat{A}^{\text{1--loop}}_{N+1}(\pmb s,\pmb n,\rho;\tau)=
\int_{0\leq z_{\rho(1)}\leq \cdots \leq z_{\rho(N)}\leq \tau}\prod_{i=1}^N\frac{dz_i}{\tau_2}\prod_{1\leq i<j\leq N+1} e^{2\alpha's_{ij} G^{\mathcal A}(z_i,z_j;\tau)}\left(\partial_{z_i} G^{\mathcal A}(z_i,z_j;\tau)\right)^{-2n_{ij}},
\end{multline}
where we set $z_{N+1}:=0$. The Green function $G^{\mathcal A}(z,z';\tau)$ from~(\ref{e:Gann}) reduces for $z,z'\in \tau\mathbb{R}/\tau\mathbb{Z}$ and $\tau\in i\mathbb{R}^+$ to 
\begin{equation}\label{Gannopen}
G^{\mathcal A}(z,z';\tau)=-2\log\left|\theta_1(z-z',\tau)  \over
    \eta(\tau)\right|-{2\pi i (z-z')^2\over \tau}\,.
\end{equation}

It was shown in~\cite{Broedel:2014vla} that the coefficients of the small $\alpha'$ expansion of the integrals~(\ref{partamplopenone}) (or rather their analytic continuation for $\tau\in\mathbb{H}$) can be written in terms of \emph{elliptic multiple zeta values}. The latter are functions on the complex upper half-plane which constitute the natural genus-one generalization of multiple zeta values and arise from the genus-one Knizhnik-Zamolodchikov-Bernard connection~\cite{Enriquez}. In particular, in the limit $\tau\rightarrow i\infty$ one lands on the space of classical multiple zeta values. This is consistent with the fact that the limit of the integrals~(\ref{partamplopenone}) can be expressed in terms of $(N+3)$-point tree-level building blocks~\cite{Mafra:2019ddf}.

From the cohomological viewpoint, elliptic multiple zeta values are periods of the configuration space of points on the torus, and it is conjectured that modular graph functions and forms belong to the family constituting their single-valued analogues~\cite{BrownNewClass}. This would be consistent with the genus-zero situation, and also with the conjecture that the coefficients of the constant terms of modular graph functions are single-valued multiple zeta values. Some evidence of an extension of the genus-zero single-valued projection between open and closed string building blocks was found in~\cite{Broedel:2018izr, Gerken:2018jrq, PanzerTalk, GKS2}. In the next section we report on one of the conjectures related to this single-valued projection, which was recently proven in the two-point case~\cite{Zagier:2019eus}.

\kern-.3cm
\subsubsection{The two-point case}

We consider the open string integral
\begin{equation}\label{partamplopenone2}
\int_{z\in [0,\tau]}e^{s\,\hat{G}^{\mathcal A}(z,0;\tau)}\,\frac{dz}{\tau_2}.
\end{equation}
This integral coincides with the building block $\hat{A}^{\text{1--loop}}_{2}(\pmb s,\pmb 0,\text{Id};\tau)$ upon setting $s:=2\alpha's_{12}$ and replacing the Green function $G^{\mathcal A}(z,z';\tau)$ from~(\ref{Gannopen}) with the modified version introduced in~\cite{Broedel:2018izr}
\begin{equation}
\hat{G}^{\mathcal A}(z,z';\tau)\,:=\,G^{\mathcal A}(z,z';\tau)\,-\,\frac{\pi i}{3\tau}\,,
\end{equation} 
which satisfies the condition $\int_{[0,\tau]}\hat{G}^{\mathcal A}(z,0;\tau)dz=0$. The correction to the Green function is constant in~$z$, so it disappears by momentum conservation when we consider the two-point amplitude (which is not physically meaningful) as contributing to higher-point amplitudes.

The integral in~(\ref{partamplopenone2}) defines a holomorphic function of~$s$ at the origin, whose $n$-th Taylor coefficient at $s=0$ (i.e. in the limit $\alpha'\rightarrow 0$) is given up to a constant factor by the integral
\begin{equation}
B_n(\tau)\,:=\,\frac{1}{4^n}\int_{z\in [0,\tau]}\,\hat{G}^{\mathcal A}(z,0;\tau)^n\,\frac{dz}{\tau_2},
\end{equation}
which can be analytically continued to define a holomorphic function of $\tau\in\mathbb{H}$. These Taylor coefficients are special cases of a class of functions arising from symmetrized open string integrals, called \emph{holomorphic graph functions}, which constitute holomorphic analogues of modular graph functions~\cite{Broedel:2018izr, Zerbini:2018hgs}. More precisely, the functions $B_n(\tau)$ are exactly those holomorphic graph functions associated to the two-points banana graphs, and so they are the open-string analogues of the two-point modular graph functions $D_n(\tau)$ from section~\ref{sec:two-points-function}.

It was proven in~\cite{Zerbini:2018hgs} that all holomorphic graph functions (associated to a graph $\Gamma$ with $E$ edges) have the asymptotic expansion
\begin{equation}
\sum_{k=-E}^E \sum_{m\geq 0} b_k^{(m)}(\Gamma)\,T^k q^m,
\end{equation}
where $T:=\pi \tau/2i$ and $b_k^{(m)}$ belong to the ring $\mathcal{Z}$ of multiple zeta values. This expansion is similar to the asymptotic expansion~(\ref{asympmodgr}) of modular graph functions, and it was conjectured in~\cite{Broedel:2018izr} that the coefficients $b_k^{(0)}(\Gamma)$ of the leading Laurent polynomial are mapped precisely to the corresponding coefficients $d_k^{(0,0)}(\Gamma)$ via the single-valued projection on multiple zeta values. 

This conjecture was recently proven for the two-point banana graphs in~\cite{Zagier:2019eus}. In other words, one can map the asymptotic limit of the two-point open string integral~(\ref{partamplopenone2}) to the asymptotic limit of the two-point closed string integral~(\ref{hhhh}) using the single-valued projection on multiple zeta values. Moreover, these asymptotic limits were also shown to be related by some double-copy formula~\cite{Zagier:2019eus}. This is not so surprising, if we think that the tree-level KLT double-copy relations can be seen as part of the same cohomological framework which explains also the single-valued projection~\cite{Brown:2018omk}. For this reason, relating the whole asymptotic expansions via the theory of single-valued periods would probably also point towards a first genus-one analogue of the KLT relations.

\subsection{Open questions}\label{sec:open}

We have already mentioned that it is not clear in general how the building blocks of annulus, M\"obius strip and Klein bottle partial amplitudes relate to each others. It would be interesting to know, for instance, whether we can reduce them to the open string building blocks~(\ref{partamplopenone}) and thus obtain the same kind of structure which occurs at tree-level. One could also ask the (less ambitious) question of what is the minimal set of special functions which appear in the $\alpha'$-expansion of these partial amplitudes: are elliptic multiple zeta values sufficient?

Moreover, at tree-level we have a clear picture not only of the building blocks, but also of the relations among them, namely the KLT relations and the single-valued projection on the $\alpha'$-expansion. An important open problem is to generalize these relations to genus one, at least for the open and closed partial amplitude building blocks~(\ref{e:Mpartial})
and~(\ref{partamplopenone}). 

For the KLT relations, one possible approach is the one presented at
tree-level in section~\ref{sec:conf-block-decomp}, based on viewing
closed string building blocks as special values of conformal
correlation functions, which have a holomorphic factorisation on
conformal blocks also at higher
genus~\cite{DiFrancesco:1997nk,DHoker:1989cxq}. At
tree-level, this conformal block factorisation gives back the KLT
relations, whence come the hope for generalized KLT relations, in
particular thanks to the chiral splitting of the closed string
integrand on a product of open string integrands at fixed loop momenta~\cite{Mafra:2018qqe}. Another possible approach is based on generalizing the interpretation of the KLT relations in terms of twisted de Rham
cohomology~\cite{Mizera:2017cqs, Brown:2018omk} to
higher-loop order~\cite{Tourkine:2016bak,Hohenegger:2017kqy,Casali:2019ihm,Casali:2020knc}.

A related open problem is to prove the conjectures which connect genus-one open and closed building blocks via the single-valued period formalism. A simpler version of this problem is to prove the conjecture that the coefficients of the constant term of any modular graph function (or form) are single-valued multiple zeta values. The new approach presented in section~\ref{sec:two-points-function} for the two-point case has the potential to be generalized for any number of points, because it is based on the same techniques used in the $N$-point case at genus zero. One must, however, overcome certain technical difficulties which arise already in the three-point case.

Finally, it would also be interesting at one-loop level to study the moduli space integrals which give the full amplitude building blocks. While some results are known in the closed oriented string case~\cite{Green:2008uj, DHoker:2019mib, DHoker:2019blr}, as far as we can tell nothing is known for the other cases. In particular, it would be interesting to understand whether we should expect to generalize in some way the KLT relations and the single-valued projection to the genus-one moduli space integrals.

\newpage
%%%%%%%%%%%%%%%%%%%%%%%%%%%%%%%%%%%%%%%%%%%%%%%%%%%%%%%%%%%%%%%%%
\bibliographystyle{ieeetr}
%\nocite{*}
%\clearpage
%\addcontentsline{toc}{chapter}{Bibliografia}
% \bibliography{biblio}
\markboth{\textsc{Bibliography}}{\textsc{Bibliography}}

\end{document}